\documentclass[aps,prc,twocolumn,superscriptaddress,showpacs]{revtex4-1}
\usepackage{graphicx}
\usepackage{amsmath} 
\usepackage[T1]{fontenc}
\usepackage{multirow}
\usepackage{tabularx}
\usepackage{comment}
\usepackage{xcolor}
\usepackage{threeparttable}
\usepackage{bm}
\usepackage{lipsum}
\usepackage{setspace}
\usepackage{braket}
\usepackage[normalem]{ulem}
\usepackage[colorlinks=true,allcolors=blue]{hyperref}

\begin{document}

\title{Probing the Size of Neutron and Proton Single-Particle Orbitals \\ from Nucleon Knockout Reactions}

\newcommand{\aatomki}{     \affiliation{HUN-REN Institute for Nuclear Research, HUN-REN ATOMKI, P.O. Box 51, Debrecen H-4001, Hungary}}
\newcommand{\abeijing}{    \affiliation{State Key Laboratory of Nuclear Physics and Technology, Peking University, Beijing 100871, China}}
\newcommand{\abnu}{\affiliation {Key Laboratory of Beam Technology and Material Modification of Ministry of Education, College of Nuclear Science and Technology, Beijing Normal University, Beijing 100875, China}}
\newcommand{\acaen}{       \affiliation{LPC Caen, Normandie Univ, ENSICAEN, UNICAEN, CNRS/IN2P3, F-14000 Caen, France}}
\newcommand{\acea}{        \affiliation{IRFU, CEA, Universit\'e Paris-Saclay, F-91191 Gif-sur-Yvette, France}}
\newcommand{\acns}{        \affiliation{Center for Nuclear Study, University of Tokyo, RIKEN campus, Wako, Saitama 351-0198, Japan}}
\newcommand{\aewha}{       \affiliation{Ewha Womans University, Seoul 03760, Korea}}
\newcommand{\aibs}{        \affiliation{Institute for Basic Science, Daejeon 34126, Korea}}
\newcommand{\agsi}{        \affiliation{GSI Helmholtzzentrum f\"ur Schwerionenforschung GmbH, Planckstr. 1, 64291 Darmstadt, Germany}}
\newcommand{\ahku}{        \affiliation{Department of Physics, The University of Hong Kong, Pokfulam, Hong Kong}}
\newcommand{\ainst}{       \affiliation{Institute for Nuclear Science \& Technology, VINATOM, 179 Hoang Quoc Viet, Cau Giay, Hanoi, Vietnam}}
\newcommand{\aipno}{       \affiliation{IPN Orsay, CNRS, Univ. Paris Sud, Univ. Paris-Saclay, F-91406 Orsay Cedex, France}}
\newcommand{\aijclab}{     \affiliation{Universit\'e Paris-Saclay, CNRS/IN2P3, IJCLab, F-91405 Orsay cedex, France}}
\newcommand{\aithems}{     \affiliation{RIKEN Center for Interdisciplinary Theoretical and Mathematical Sciences (iTHEMS), RIKEN, Wako 351-0198, Japan}}
\newcommand{\akoeln}{      \affiliation{Institut f\"ur Kernphysik, Universit\"at zu K\"oln, D-50937 Cologne, Germany}}
\newcommand{\akth}{        \affiliation{Department of Physics, Royal Institute of Technology, SE-10691 Stockholm, Sweden}}
\newcommand{\alanzhou}{    \affiliation{Institute of Modern Physics, Chinese Academy of Sciences, Lanzhou 730000, China}}
\newcommand{\amadrid}{     \affiliation{Instituto de Estructura de la Materia, CSIC, E-28006 Madrid, Spain}}
\newcommand{\amsu}{        \affiliation{Department of Physics and Astronomy, Michigan State University, East Lansing, MI 48824-1321, United States}}
\newcommand{\aorsay}{      \affiliation{CSNSM, CNRS/IN2P3, Universit\'e Paris-Sud, F-91405 Orsay Campus, France}}
\newcommand{\aoslo}{       \affiliation{Department of Physics, University of Oslo, N-0316 Oslo, Norway}}
\newcommand{\arcnp}{      \affiliation{Research Center for Nuclear Physics (RCNP), Osaka University, Ibaraki 567-0047, Japan}}
\newcommand{\ariken}{      \affiliation{RIKEN Nishina Center, 2-1 Hirosawa, Wako, Saitama 351-0198, Japan}}
\newcommand{\arikkyo}{     \affiliation{Department of Physics, Rikkyo University, 3-34-1 Nishi-Ikebukuro, Toshima, Tokyo 172-8501, Japan}}
\newcommand{\atitech}{     \affiliation{Department of Physics, Institute of Science Tokyo, 2-12-1 O-Okayama, Meguro, Tokyo, 152-8551, Japan}}
\newcommand{\atohoku}{     \affiliation{Department of Physics, Tohoku University, Sendai 980-8578, Japan}}
\newcommand{\atudarmstadt}{\affiliation{Institut f\"ur Kernphysik, Technische Universit\"at Darmstadt, 64289 Darmstadt, Germany}}
\newcommand{\aunal}{       \affiliation{Universidad Nacional de Colombia, Sede Bogot\'a, Facultad de Ciencias, Departamento de Física, Bogot\'a 111321, Colombia}}
\newcommand{\ajaver}{      \affiliation{Pontificia Universidad Javeriana, Facultad de Ciencias, Departamento de F\'isica, Bogot\'a, Colombia}}
\newcommand{\aut}{         \affiliation{Department of Physics, University of Tokyo, 7-3-1 Hongo, Bunkyo, Tokyo 113-0033, Japan}}
\newcommand{\azagreb}{     \affiliation{Ru{\dj}er Bo\v{s}kovi\'c Institute, Bijeni\v{c}ka cesta 54, 10000 Zagreb, Croatia}}
\newcommand{\abro}{        \affiliation{Laboratoire Kastler Brossel, Sorbonne Universit\'e, CNRS, ENS, PSL Research University, Coll\`ege de France, Case 74, 4 Place Jussieu, 75005 Paris, France}}
\newcommand{\APoves}{        \affiliation{Departamento de Fisica Teorica and IFT UAM-CSIC, Universidad Autonoma de Madrid, Spain}}
\newcommand{\FNowacki}{        \affiliation{Université de Strasbourg, CNRS, IPHC UMR 7178, F-67000 Strasbourg, France}}
\newcommand{\KOgata}{        \arcnp }
\newcommand{\KOgataS}{        \affiliation{Department of Physics, Kyushu University, Fukuoka 819-0395, Japan}}
\newcommand{\KYoshida}{        \affiliation{Advanced Science Research Center, Japan Atomic Energy Agency, Tokai, Ibaraki 319-1195, Japan}}
\newcommand{\afair}{      \affiliation{Helmholtz Forschungsakademie Hessen für FAIR (HFHF), GSI Helmholtzzentrum für Schwerionenforschung, Campus Darmstadt, 64289 Darmstadt, Germany }}
\newcommand{\EMMI}{
\affiliation{ExtreMe Matter Institute EMMI, GSI Helmholtzzentrum f\"ur Schwerionenforschung GmbH, 64291 Darmstadt, Germany}}
\newcommand{\MaxPlanckHeidelberg}{\affiliation{Max-Planck-Institut f\"ur Kernphysik, Saupfercheckweg 1, 69117 Heidelberg, Germany}}
\newcommand{\takayukinew}{\affiliation{Center for Computational Sciences, University of Tsukuba, 1-1-1 Tennodai, Tsukuba 305-8577, Japan}}
\newcommand{\milano}{\affiliation{Dipartimento di Fisica “Aldo Pontremoli”, Universit\`a degli Studi di Milano, via Celoria 16, I-20133 Milano, Italy}}
\newcommand{\infn}{\affiliation{INFN, Sezione di Milano, via Celoria 16, I-20133 Milano, Italy}}
\newcommand{\lanzhou}{\affiliation{School of Nuclear Science and Technology, Lanzhou University}}

\newcommand{\aethzipp}{\affiliation{Institute of Particle Physics and Astrophysics, ETH Z\"urich, CH-8093 Z\"urich, Switzerland}}
\newcommand{\aornl}{\affiliation{National Center for Computational Sciences, Oak Ridge National Laboratory, Oak Ridge, Tennessee 37831, USA}}
\newcommand{\aomu}{\affiliation{Department of Physics, Osaka Metropolitan University, Osaka 558-8585, Japan}}
\newcommand{\aomuNambu}{\affiliation{Nambu Yoichiro Institute of Theoretical and Experimental Physics, Osaka Metropolitan University, Osaka 558-8585, Japan}}
\newcommand{\aiststrategy}{\affiliation{Office of Institute Strategy, Institute of Science Tokyo, Tokyo 152-8550, Japan}}

\author{M.~Enciu} \email{menciu@ethz.ch} \atudarmstadt \aethzipp
\author{A.~Obertelli} \atudarmstadt \acea \ariken 
\author{P.~Doornenbal} \ariken
\author{C. Barbieri} \milano \infn
\author{S. Brolli} \milano \infn
\author{M. Heinz}\atudarmstadt \EMMI \MaxPlanckHeidelberg \aornl
\author{W. Horiuchi} \aomu \aomuNambu \ariken
\author{T. Inakura} \aiststrategy
\author{W. H. Long} \lanzhou
\author{T. Miyagi}\atudarmstadt \EMMI \MaxPlanckHeidelberg \takayukinew
\author{F.~Nowacki} \FNowacki
\author{K.~Ogata}\KOgataS \KOgata
\author{A.~Poves}\APoves
\author{A. Schwenk} \atudarmstadt \EMMI \MaxPlanckHeidelberg
\author{K.~Yoshida}\KYoshida \arcnp \aithems
\author{N.~L.~Achouri}\acaen
\author{H.~Baba} \ariken
\author{F.~Browne}\ariken
\author{D.~Calvet} \acea
\author{F.~Ch\^ateau} \acea
\author{S.~Chen} \ahku \ariken \abeijing          
\author{N.~Chiga}\ariken 
\author{A.~Corsi} \acea 
\author{M.~L.~Cort\'es} \ariken
\author{A.~Delbart} \acea
\author{J-M.~Gheller} \acea
\author{A.~Giganon}\acea                          
\author{A.~Gillibert} \acea
\author{C.~Hilaire} \acea
\author{T.~Isobe} \ariken 
\author{T.~Kobayashi}\atohoku
\author{Y.~Kubota} \ariken \acns
\author{V.~Lapoux} \acea
\author{H.~N.~Liu} \atudarmstadt \acea \akth  
\author{T.~Motobayashi} \ariken                 
\author{I.~Murray} \aijclab \ariken
\author{H.~Otsu} \ariken
\author{V.~Panin}\ariken
\author{N.~Paul} \acea \abro                    
\author{W.~Rodriguez} \ariken \ajaver \aunal
\author{H.~Sakurai} \ariken \aut
\author{M.~Sasano} \ariken
\author{D.~Steppenbeck}\ariken
\author{L.~Stuhl}\acns \aatomki \aibs             
\author{Y.~L.~Sun}  \acea \atudarmstadt       
\author{Y.~Togano}\arikkyo \ariken 
\author{T.~Uesaka} \ariken
\author{K.~Wimmer}\aut \ariken                    
\author{K.~Yoneda} \ariken 
\author{O.~Aktas}\akth
\author{T.~Aumann}\atudarmstadt \agsi             
\author{L.~X.~Chung} \ainst 
\author{F.~Flavigny}\aijclab  \acaen
\author{S.~Franchoo}\aijclab 
\author{I.~Gasparic}\azagreb \atudarmstadt \ariken
\author{R.-B.~Gerst}\akoeln
\author{J.~Gibelin}\acaen
\author{K.~I.~Hahn} \aewha \aibs
\author{D.~Kim} \aewha \ariken \aibs
\author{Y.~Kondo}\atitech                         
\author{P.~Koseoglou}\atudarmstadt \agsi
\author{J.~Lee} \ahku
\author{C.~Lehr}\atudarmstadt                     
\author{P.~J.~Li} \ahku
\author{B.~D.~Linh} \ainst 
\author{T.~Lokotko}\ahku
\author{M.~MacCormick}\aijclab
\author{K.~Moschner}\akoeln
\author{T.~Nakamura}\atitech                   
\author{S.~Y.~Park} \aewha \aibs 
\author{D.~Rossi}\atudarmstadt                  
\author{E.~Sahin} \aoslo
\author{P-A.~S\"oderstr\"om} \atudarmstadt
\author{D.~Sohler}\aatomki  
\author{S.~Takeuchi} \atitech
\author{H.~Toernqvist}\atudarmstadt  \agsi        
\author{V.~Vaquero}\amadrid 
\author{V.~Wagner}\atudarmstadt                   
\author{S.~Wang}\alanzhou 
\author{V.~Werner}\atudarmstadt
\author{X.~Xu} \ahku
\author{H.~Yamada}\atitech                       
\author{D.~Yan} \alanzhou
\author{Z.~Yang} \ariken
\author{M.~Yasuda}\atitech                       
\author{L.~Zanetti}\atudarmstadt

\begin{abstract}
The size of neutron and proton single-particle orbitals of $^{52}$Ca, $^{53}$Ca, $^{54}$Ca, and $^{55}$Sc were investigated via nucleon knockout reactions at $\sim$ 230\,MeV/nucleon. The determination method is based on the measured fragment momentum distributions in $(p,pn)$ and $(p,2p)$ reactions, which are shown to be sensitive to the spatial extension of the wave function of the knocked-out nucleon, interpreted within the distorted wave impulse approximation (DWIA) framework. A systematic sensitivity study is carried out for the $(p,pn)$ recoil-momentum distribution method and is presented in this work. The experimental momentum distributions are compared to state-of-the-art mean field and {\it ab initio} in-medium similarity renormalization group and self-consistent Green's function calculations in combination with DWIA reaction theory calculations. Based on this work, the 1$p$ neutron orbitals are consistently found $0.48-0.78$~fm larger than the $0f_{7/2}$ neutron orbitals in $^{52-54}$Ca, while the size evolution of the valence proton orbitals remains inconclusive due to the large associated statistical uncertainties.
\end{abstract}
\maketitle

\section{INTRODUCTION}
Understanding the size of atomic nuclei—specifically the matter, charge, and neutron radii—is essential to unraveling the complex interplay of forces within nuclear matter~\cite{Hagen:2015yea}. The charge radii of both stable and unstable nuclei have been measured with remarkable precision via optical probes~\cite{Nortershauser2023,nudat_chargeradii}. Neutron and matter radii are measured model-dependently by elastic scattering with hadronic and leptonic probes as well as via interaction cross section measurements, see~\cite{Zenihiro2010,Rossi2013,Adhikari2022,Tanaka2020}. Most of these methods are limited to stable nuclei and thus the size of neutron radii and the neutron skin thickness information is missing for most unstable isotopes, with the few measured cases lacking high precision.

Neutron and proton quasi-free scattering is sensitive to single-particle properties of nuclei as opposed to bulk properties~\cite{Aumann2021}. The size of valence proton and neutron orbitals can be probed by one-nucleon knockout reactions~\cite{Enciu2022}. Probing the size of single-particle orbitals has been studied with knockout, transfer, and $(e,e'p)$ reactions~\cite{Enciu2022,Durell1980,denherder1988,Kramer1989,Kramer2001}. The size of single-particle orbitals and the isotope shifts of single-particle rms radii~\cite{Friedmann1977} can give information about relative changes between neighboring nuclei and helps in tracking the effect of one nucleon type on the opposite type of nucleons in nuclei. The calcium isotopes provide an ideal scenario with $Z=20$ and $N$ spanning the $sd$ and the $pf$ shells. The charge radii in the Ca isotopes were found to increase steeply after $N=28$ while state-of-the-art microscopic calculations could not reproduce the experimental measurements~\cite{GarciaRuiz2016,Heinz2024,Lellinger2026}. This behavior is not limited to the Ca isotopic chain, but extends to its neighbors, which shows systematic and sizable charge radii increase as the $1p$ neutron orbitals are filled~\cite{nudat_chargeradii,Enciu2022,Bonnard2016,Koszorus2021,Sommer2022}. Moreover, a core swelling effect being triggered by the filling of large $1p$ neutron orbitals has been proposed~\cite{Tanaka2020,Horiuchi2020,Liu2020}.
In this paper, we present a study of the size of the 0$f_{7/2}$ and $1p$ neutron single-particle orbitals of $^{53,54}$Ca, of the $1s_{1/2}$ and $0d_{3/2}$ proton orbitals of $^{52-54}$Ca, and of the $0f_{7/2}$ proton orbital of $^{55}$Sc determined from exclusive recoil-momentum distributions of quasi-free scattering reactions. The results give insights into the matter and charge radii behavior as neutrons are added to the $^{48}$Ca core. A detailed sensitivity study is performed for validating our method for extracting single-particle root-mean-square (rms) radii via momentum distributions. The results are compared to \emph{ab initio} in-medium similarity renormalization group (IMSRG) and self-consistent Green's function (SCGF) calculations based on chiral effective field theory interactions, as well as to mean-field calculations using different Skyrme energy-density functionals.

\begin{table*}[t]
\caption{Final nucleus and the final state (excitation energy $E_{\mathrm{x}}$ and spin-parity $J^\pi$) for each case of knocking out a proton or a neutron from the initial $^{52,53,54}$Ca and $^{55}$Sc nuclei. Fifth and sixth columns list the experimental exclusive and inclusive cross sections, $\sigma^{\mathrm{exp}}_{\mathrm{exc}}$ and $\sigma^{\mathrm{exp}}_{\mathrm{incl}}$. The last three columns list the single-particle orbital of the knocked-out nucleon, the optimum radial parameter, r$_0$, and the single-particle orbital size, R$_{\mathrm{s.p.}}$, determined in this study (see text for details).}
\label{table_rms_radii}
\begin{spacing}{1.1}
\centering
\begin{threeparttable}
\begin{tabular}{|>{\centering\arraybackslash}p{2cm} |>{\centering\arraybackslash}p{1.5cm} >{\centering\arraybackslash}p{1.5cm} >{\centering\arraybackslash}p{1.5cm}|>{\centering\arraybackslash}p{2cm} >{\centering\arraybackslash}p{2cm}|>{\centering\arraybackslash}p{2cm} >{\centering\arraybackslash}p{2cm} >{\centering\arraybackslash}p{2cm}|}
\hline
\begin{tabular}[c]{c}Initial\\ nucleus\end{tabular} & \begin{tabular}[c]{c}Final\\ nucleus\end{tabular} & \begin{tabular}[c]{c}$E_{\mathrm{x}}$\\ (keV)\end{tabular} & \begin{tabular}[c]{c}$J^{\pi}$\end{tabular} & \begin{tabular}[c]{c}$\sigma^{\mathrm{exp}}_{\mathrm{exc}}$\\ (mb)\end{tabular} & \begin{tabular}[c]{c}$\sigma^{\mathrm{exp}}_{\mathrm{incl}}$ \\ (mb)\end{tabular} & \begin{tabular}[c]{c}single-part.\\ orbital\end{tabular} & \begin{tabular}[c]{c}r$_{0}$\\ (fm)\end{tabular} & \begin{tabular}[c]{c}R$_{\mathrm{s.p.}}$\\ (fm)\end{tabular}\\ \hline
\hline
$^{52}$Ca & $^{51}$K & g.s. & $3/2^+$ & 5.65(67) & \multirow{2}{*}{9.00(57)$^{(a)}$} & $\pi 0d_{3/2}$ & 1.48(9) & 3.98(18) \\    
$^{52}$Ca & $^{51}$K & 737 & $1/2^+$ & 1.48(23) & \multicolumn{1}{c|}{} & $\pi 1s_{1/2}$ & 1.32(21) & 3.57(33) \\ \hline
\hline
$^{53}$Ca & $^{52}$K & g.s. & $2^-$ & 3.21(43) & \multirow{2}{*}{5.84(39)$^{(b)}$} & $\pi 0d_{3/2}$ & 1.55(7) & 4.11(14) \\
$^{53}$Ca & $^{52}$K & 1076 & $1^-_2$ & 0.81(12) & \multicolumn{1}{c|}{} & $\pi 1s_{1/2}$ & 1.42(16) & 3.72(26) \\
\hline
$^{53}$Ca & $^{52}$Ca & g.s. & $0^+$ & 15.1(24) & \multirow{4}{*}{38.5(13)} & $\nu 1p_{1/2}$ & 1.17(13) & 4.86(21) \\ 
$^{53}$Ca & $^{52}$Ca & 2563 & $2^+$ & 9.9(16) & \multicolumn{1}{c|}{} & $\nu 1p_{3/2}$ & 1.29(6) & 4.67(10) \\ 
$^{53}$Ca & $^{52}$Ca & 3150 & $1^+$ & 5.3(9) & \multicolumn{1}{c|}{} & $\nu 1p_{3/2}$ & 1.30(7) & 4.61(13) \\ 
$^{53}$Ca & $^{52}$Ca & 5940 & ($3^+,4^+$) & 8.2(12) & \multicolumn{1}{c|}{} & $\nu 0f_{7/2}$ & 1.19(5) & 4.10(13) \\ \hline
\hline
$^{54}$Ca & $^{53}$K & g.s. & $3/2^+$ & 3.9(4)$^{(a)}$ & \multirow{2}{*}{5.2(3)$^{(a)}$} & $\pi 0d_{3/2}$ & 1.41(8)$^{(a)}$ & 3.85(17) \\
$^{54}$Ca & $^{53}$K & 837 & $1/2^+$ & 1.3(2) & \multicolumn{1}{c|}{} & $\pi 1s_{1/2}$ & 1.72(18) & 4.25(31) \\ 
\hline
$^{54}$Ca & $^{53}$Ca & g.s. & $1/2^-$ & 15.9(17) & \multirow{2}{*}{35.0(21)$^{(c)}$} & $\nu 1p_{1/2}$ & 1.25(14) & 4.88(24) \\ 
$^{54}$Ca & $^{53}$Ca & 2220 & $3/2^-$ & 19.1(12) & \multicolumn{1}{c|}{} & $\nu 1p_{3/2}$ & 1.28(14) & 4.64(24) \\ \hline
\hline
$^{55}$Sc & $^{54}$Ca & g.s. & $0^+$ & -- & -- & $\pi 0f_{7/2}$ & 1.26(11) & 4.07(26) \\ \hline
\end{tabular}
\begin{tablenotes}[flushleft]
\footnotesize
\item $^{(a)}$ Exclusive and inclusive cross sections re-evaluated based on data of Ref.~\cite{Sun}.
\item $^{(b)}$ Exclusive and inclusive cross sections as reported in Ref.~\cite{Enciu2024}.
\item $^{(c)}$ Exclusive and inclusive cross sections re-evaluated based on data of Ref.~\cite{Chen2019}.
\end{tablenotes}
\end{threeparttable}
\end{spacing}

\end{table*}
\section{EXPERIMENTAL SETUP}
\begin{figure*}[t]
    \centering
    \includegraphics[width=1\textwidth]{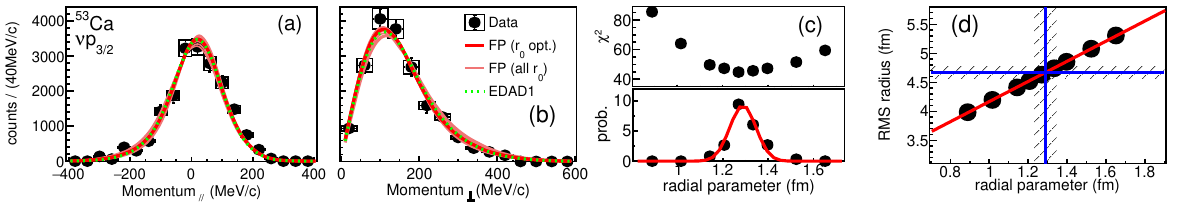}
    \includegraphics[width=1\textwidth]{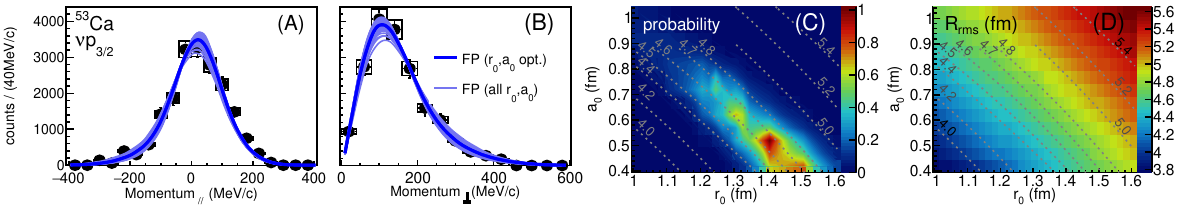}
    \caption{Parallel (panels {\it a} and {\it A}) and perpendicular (panels {\it b} and {\it B}) momentum distributions for $^{53}$Ca$(p,pn)$$^{52}$Ca ($E_{\mathrm{x}}=2.5$~MeV). The calculated momentum distributions curves are shown for all possible values of r$_0$ (top panels), respectively for all (r$_0$, a$_0$) combinations (bottom panels) as indicated in the legends. The $\chi^2$ and probability distributions as a function of the radial parameter are shown in panel {\it c} for the one-dimensional case, respectively in panel {\it C} for the two-dimensional (r$_0$, a$_0$) case. Panel {\it d} shows the correspondence of the rms radii to the radial parameter, and panel {\it D} shows the rms radii values for each (r$_0$, a$_0$) pair for the 2D case.}
    \label{pmd_53Ca_np3}
\end{figure*}

The experiment was carried out at the Radioactive Isotope Beam Factory of RIKEN, operated jointly by the RIKEN Nishina Center and the Center for Nuclear Study, University of Tokyo. A 240-pnA $^{70}$Zn primary beam at 345~MeV/nucleon bombarded a 10-mm-thick Be target for the secondary beam production. The beam identification was done event-by-event in the BigRIPS separator~\cite{BigRIPS-T.Kubo, BigRIPS-N.Fukuda} via magnetic rigidity (B$\rho$), time-of-flight (ToF) and energy-loss ($\Delta$E) measurements. The $^{52}$Ca, $^{53}$Ca, $^{54}$Ca, and $^{55}$Sc beam particles were obtained with average intensities of 4.4, 12.6, 7.1, and 4.5 particles per second, respectively, and the beam time had a total duration of 7 days. The fragments were identified and separated after knockout reactions using the large-acceptance SAMURAI magnet~\cite{SAMURAI-T.Kobayashi} and the standard SAMURAI detectors: the fragment drift chambers (FDCs) and the 24 plastic scintillator-bar hodoscope detectors by the B$\rho$-ToF-$\Delta$E method, event-by-event.
    The MINOS~\cite{MINOS-A.Obertelli, MINOS-C.Santamaria} setup, comprising a 151-mm-long liquid hydrogen target and a time projection chamber (TPC), was used for inducing the proton and neutron knockout reactions and for reaction vertex reconstruction. The reaction vertex reconstruction was obtained with a resolution of 5~mm (FWHM) by using the trajectories of a proton and the incoming beam for $(p,pn)$ reactions and the trajectories of the two protons in the case of $(p,2p)$ reactions. The incident energy at the reaction vertex position was between 160~MeV/nucleon and 280~MeV/nucleon. 
The momentum distributions presented throughout this study were obtained as the momentum of the fragments in the center of mass of the beam measured after $(p,pn)$ and $(p,2p)$ reactions. For this, the measurements were kinematically complete: the momentum of the beam was determined by the velocity measured by TOF in BigRIPS and the direction was obtained using the position information of the beam drift chamber, located before the liquid hydrogen target, and the reconstructed vertex position; the momentum of the fragment was determined by the velocity obtained using the TOF spent over the reconstructed flight length in the SAMURAI spectrometer and the direction of the fragment was obtained from the position of the reaction vertex and the position in the fragment drift chamber, positioned after the liquid hydrogen target. The momentum resolution for the parallel (perpendicular) component was determined from the unreacted events and had the following values: 49.5 (76.5), 47.8 (77.6), 52.5 (78.5), and 51.6 (80.6)~MeV/c for $^{52}$Ca, $^{53}$Ca, $^{54}$Ca, and $^{55}$Sc, respectively, given as Gaussian standard deviation ($\sigma$). The final states of the fragment were tagged via $\gamma$-ray spectroscopy using the DALI2$^+$~\cite{DALI-S.Takeuchi} high-efficiency array of 226 NaI(Tl) scintillation detectors surrounding MINOS. The exclusive momentum distributions to excited final states were obtained based on the $\gamma$-ray spectra corresponding to each momentum interval. The $\gamma$-ray spectra were fitted with the response functions of DALI2$^+$, simulated using Geant4~\cite{geant4-S.Agostinelli}, as described in Refs.~\cite{Enciu2022,Enciu2024}. 
    The reaction channels of interest for this work are $^{53}$Ca$(p,pn)^{52}$Ca, $^{54}$Ca$(p,pn)^{53}$Ca, $^{52}$Ca$(p,2p)^{51}$K, $^{53}$Ca$(p,2p)^{52}$K, $^{54}$Ca$(p,2p)^{53}$K, and $^{55}$Sc$(p,2p)^{54}$Ca. The $\gamma$-ray spectra and cross-sections for each case were reported in Refs.~\cite{Sun,Enciu2024,Chen2019,Browne2021,PJLi} using the same data set and therefore the final states populated after the knockout of the $pf$-shell neutrons and the $sd$-protons are already known. Table~\ref{table_rms_radii} lists the populated final states for each studied channel and the obtained inclusive and exclusive cross sections. The inclusive and exclusive cross sections are re-evaluated in this work, and agree very well with the reported ones in Refs.~\cite{Sun,Chen2019}. See Appendix for all momentum distribution plots for each studied reaction channel and final state discussed in this paper. We choose the $^{53}$Ca$(p,pn)$$^{52}$Ca (final state $E_{\mathrm{x}}$ = 2.5~MeV, $J^{\pi}$ = 2$^+$, $p_{3/2}$ neutron knock-out) to illustrate the method. The experimental parallel and perpendicular exclusive momentum distributions for this case are plotted in Fig.~\ref{pmd_53Ca_np3}, panels (a) and (b).
\section{Single-particle radii}
\subsection{DWIA}
For the interpretation of the experimental exclusive momentum distributions, theoretical momentum distributions were calculated with the distorted wave impulse approximation (DWIA) formalism~\cite{Chant1977,Wakasa2017,JM66,JM73,Kit85}. The folding potential (FP)~\cite{Toyokawa2013} was used with the Melbourne $G$-matrix interaction~\cite{Negele2002} for the incoming and outgoing scattering waves. The single-particle wave function of the knocked-out nucleon was obtained as a bound state of the Woods-Saxon one-body potential~\cite{Yoshida2021}, the depth of which was adjusted to match the effective separation energy of the nucleon, while the radial and diffuseness parameters were varied to fit the experimental momentum distributions. The scattering and bound-state wave functions were corrected for non-locality by using the Perey factor~\cite{Per63} and the nucleon-nucleon effective interaction parameterized by Franey and Love~\cite{Franey1985} was used to describe the elementary $p$-$n$ scattering process. For the neutron and proton distributions of the parent and daughter nuclei, Hartree-Fock-Bogoliubov (HFB) calculations with the SKM$^\ast$ interaction~\cite{Bartel1982} were performed using the \texttt{hfbrad} code~\cite{hfbrad}. Theoretical momentum distributions were calculated using the computer code {\sc pikoe}~\cite{pikoe24} and convoluted with the reaction energy profile and the experimental momentum resolution for each reaction channel.

\subsection{Extraction of single-particle rms radii}
An effective size of the single-particle orbital from which the nucleon is removed via the knock-out reaction is determined by fitting the theoretical momentum distribution curves to the experimental ones while varying the radial parameter as in Ref.~\cite{Enciu2022}. The experimental momentum resolution that we obtain is smaller than the intrinsic momentum spread, allowing the extraction of effective single-particle radii. The value for the optimum r$_{0}$ is found using a $\chi^2$ minimization criterion. Figures~\ref{pmd_53Ca_np3} (a) and (b) show the theoretical calculations for a variation of the radial parameter between 0.8 and 1.8~fm while keeping $a_0 =0.67$~fm. Figure~\ref{pmd_53Ca_np3} (c) shows the $\chi^2$ distribution ($\chi_{//}^2$+$\chi_{\perp}^2$) and the determination of the optimum radial parameter fitting a Gaussian on the probability distribution. Figure~\ref{pmd_53Ca_np3} (d) shows the relation between the single-particle orbital size and the radial parameter, based on which the single-particle orbital size ($R_{\mathrm{s.p.}}$) is determined. The uncertainty in the extracted value of $R_{\mathrm{s.p.}}$ is primarily determined by the 1-$\sigma$ uncertainty in the optimal radial parameter (r$_{0}$), which is predominantly statistical in nature. The optimum r$_{0}$ and $R_{\mathrm{s.p.}}$ obtained via this method are listed in Table~\ref{table_rms_radii} for all cases. The results for the neutron orbitals for $^{53,54}$Ca are in agreement with the findings of Ref.~\cite{Enciu2022} and the prediction of Ref.~\cite{Bonnard2016} for $^{52}$Ca, i.e., the $1p_{3/2}$ and $1p_{1/2}$ orbitals are larger than the $0f_{7/2}$ orbital by $0.48-0.78$~fm. Furthermore, the $1p_{1/2}$ are slightly larger than the $1p_{3/2}$ neutron orbitals. For the proton orbitals we observe that the uncertainties are much larger than for the neutron orbitals---this is due to the lower statistics for the $(p,2p)$ reaction channels. The size of the $0d_{3/2}$ orbitals is constant for all three Ca isotopes, while the $1s_{1/2}$ orbitals indicate a slight increase towards $^{54}$Ca.

\subsection{Sensitivity study}
To validate our single-particle radii determination method and to investigate the impact of the choice of the optical potential, additional calculations were performed using the Dirac phenomenology potential EDAD1 (Dirac)~\cite{Coo93} where the Darwin factor~\cite{Darwin_ref,Wakasa2017} was employed. After the variation of the radial parameter in the same manner for both FP and EDAD1 potentials, a difference of 0.1~fm (8.1\%) was found for the optimum radial parameter value in the case of \hbox{$^{53}$Ca$(p,pn)$$^{52}$Ca (-$\nu p_{3/2}$)} which translates into a difference of 0.16~fm (3.5\%) for $R_{\mathrm{s.p.}}$; this is considered as the uncertainty due to the optical potential (not included in the uncertainties of $R_{\mathrm{s.p.}}$ in Table~\ref{table_rms_radii}). The theoretical curve for optimum r$_{0}$ with the EDAD1 potential is shown in Fig.~\ref{pmd_53Ca_np3} (a) and (b) overlapping very well with the curves using FP.

\begin{figure}[t]
    \centering
    \includegraphics[width=0.5\textwidth]{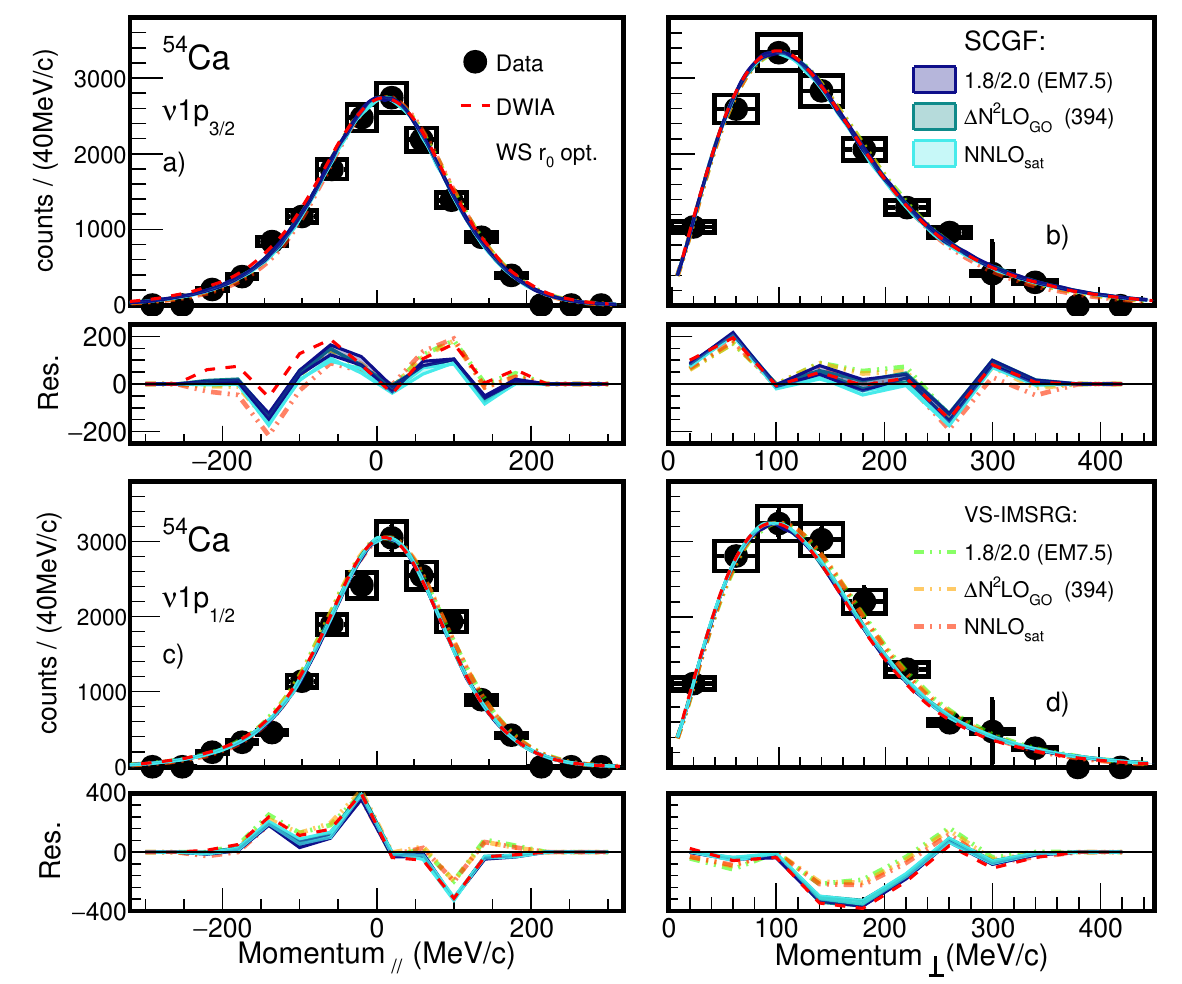}
    \caption{Exclusive momentum distribution plots for the $^{54}$Ca$\,(p,pn)\,{}^{53}$Ca reaction channel with the 1$p_{3/2}$ (panels \emph{a} and \emph{b}) and 1$p_{1/2}$ (panels \emph{c} and \emph{d}) neutron knock-out. The experimental data are shown with black filled circles, with error bars for statistical uncertainties and error boxes for systematic uncertainties. The momentum distributions are divided into the parallel (left panels) and perpendicular (right panels) projections. These are compared against theoretical calculations for the momentum distributions from DWIA calculations together with ab~initio SCGF and VS-IMSRG predictions for the transition amplitudes employing three Hamiltonians from chiral effective field theory~\cite{Ekstrom2015,Jiang2020,Arthuis2024}. Results from the SCGF method are plotted as bands indicating estimated uncertainties from model space convergence (see Appendix). For comparison, the momentum distributions computed using the DWIA and the Woods-Saxon wave functions with the optimal radial parameter are also shown with red dashed lines. To better visualize the differences between each calculated curve, each momentum distribution plot is accompanied by a panel indicating residuals (Res.) to the experimental data in units of counts per bin.}
    \label{pmdEX_54Ca_np3_np1}
\end{figure}
\begin{figure}[t]
    \centering
    \includegraphics[width=0.5\textwidth]{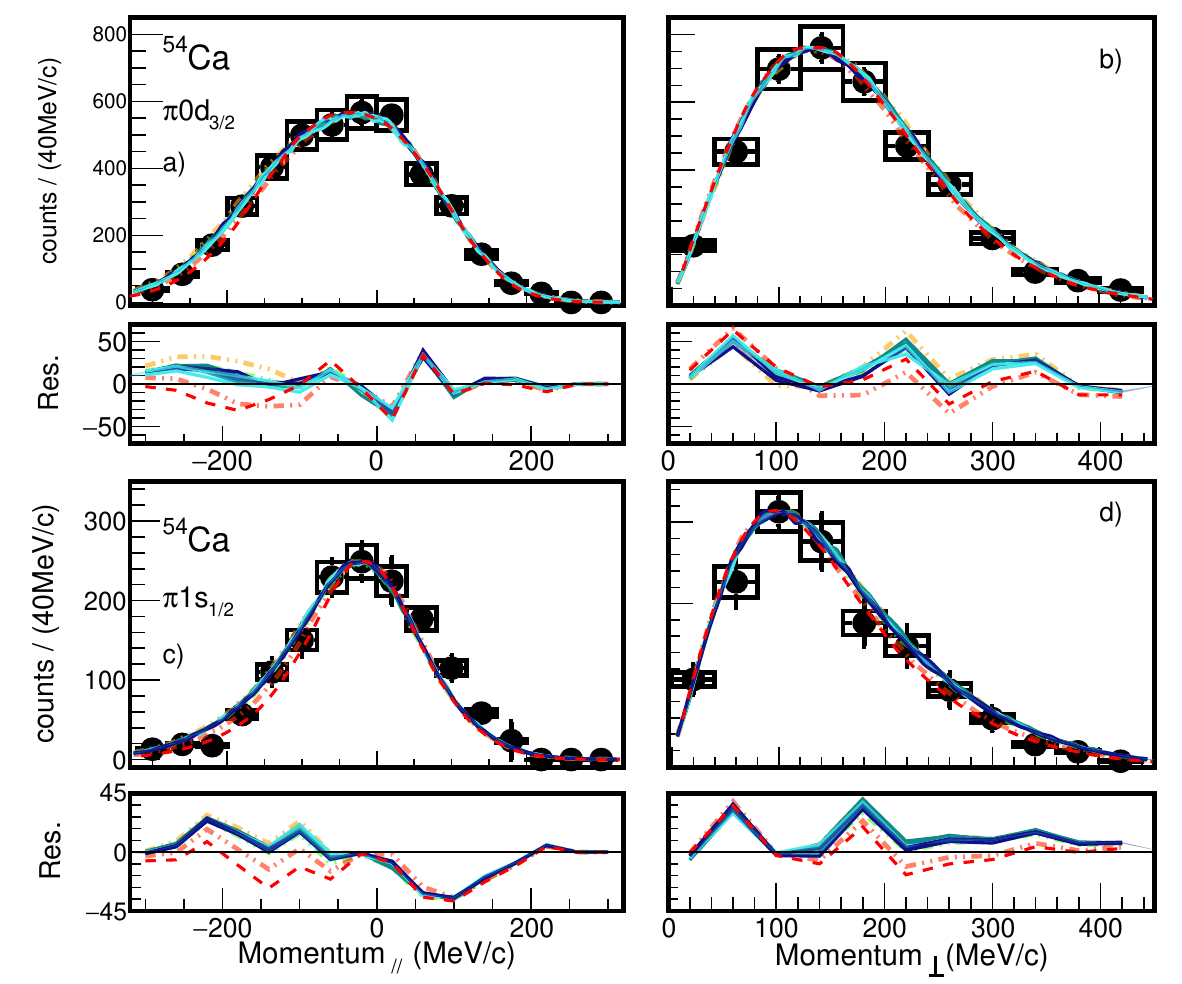}
\caption{Exclusive momentum distribution plots for the $^{54}$Ca$\,(p,2p)\,{}^{53}$K reaction channel with the 0$d_{3/2}$ (panels \emph{a} and \emph{b}) and 1$s_{1/2}$ (panels \emph{c} and \emph{d}) proton knock-out. The legend is the same as in Fig.~\ref{pmdEX_54Ca_np3_np1}.}
    \label{pmdEX_54Ca_pd3_ps1}
\end{figure}
A study for the two-dimensional (2D) variation of both the radial and the diffuseness parameters as opposed to the one-dimensional variation of r$_{0}$ was also carried out. The theoretical curves for all possible (r$_0$, a$_0$) combinations for an a$_0$ range between $0.4-1.05$~fm are shown in Fig.~\ref{pmd_53Ca_np3} (A and B). The step increment used for a$_0$ is 0.05~fm between $0.57-0.77$~fm and 0.1~fm in the rest of the scanned a$_0$ region, while the step increment used for r$_0$ is 0.01~fm between $1.236-1.316$~fm and 0.1~fm in the rest of the scanned r$_0$ region. The 2D probability distribution, in this case, is shown in Fig.~\ref{pmd_53Ca_np3} (C) as a contour plot. Additionally, Fig.~\ref{pmd_53Ca_np3} (D) plots the $R_{\mathrm{s.p.}}$ distribution as a function of (r$_0$, a$_0$) where one can see isovalue lines distributed diagonally showing that the (r$_0$, a$_0$) pairs are not uniquely corresponding to one $R_{\mathrm{s.p.}}$ value. The maximum probability in Fig.~\ref{pmd_53Ca_np3} (C) is distributed along one line in the (r$_0$, a$_0$) space corresponding to $R_{\mathrm{s.p.}}$ = 4.66(7)~fm, with the peak for (r$_0$, a$_0$)=(1.406~fm, 0.52~fm). From the 1D variation of r$_0$ only, the result of $R_{\mathrm{s.p.}}$ = 4.67(15)~fm was obtained which is in very good agreement with the result from the 2D variation. Based on these findings, we chose to continue to use only the variation of the radial parameter with fixed diffuseness parameter (a$_0$ = 0.67~fm) for our single-particle rms radii determination. The fixed diffuseness parameter of 0.67 fm was chosen based on the Bohr-Mottelson parametrization~\cite{Bohr-Mottelson}, which is typically used for stable nuclei in this mass range and has been shown to provide a good description of nuclear density distributions.

\section{Theoretical predictions}
\subsection{VS-IMSRG and DWIA}
We employ the {\it ab initio} valence-space in-medium similarity renormalization group (VS-IMSRG)~\cite{Hergert2016,Stroberg2019,Heinz2021}, which transforms an input Hamiltonian based on two- (NN) and three-nucleon (3N) interactions to decouple an effective valence-space Hamiltonian, truncated at the normal-ordered two-body level, the VS-IMSRG(2).
We generated the input NN and 3N matrix elements using the \texttt{NuHamil} code~\cite{NuHamil},
and we solved the VS-IMSRG using the \texttt{imsrg++} code~\cite{imsrg++}.
The Hamiltonian was transformed to decouple the proton $sd$-shell and neutron $pf$-shell valence space, and the valence-space problem was solved using the \texttt{kshell} code~\cite{KSHELL}. 
This valence-space Hamiltonian,
specifically computed using the initial nucleus for the reference state, is diagonalized to compute the eigenstates of the initial and final nuclei. The transition amplitude is computed based on the one-body annihilation operator $\hat{a}_{nlj}$ for the $nlj$ orbital.
The calculations are performed in the basis of 15 major harmonic oscillator shells for a frequency of $\hbar \omega = 12$~MeV, where the computed point-proton and point-neutron radii show a better convergence.
To test the nuclear Hamiltonian sensitivity, we employ three Hamiltonians from chiral effective field theory~\cite{Epelbaum2009,Machleidt2011}: the NNLO$_{\mathrm{sat}}$~\cite{Ekstrom2015}, $\Delta$N$^{2}$LO$_{\rm GO}$\,(394)~\cite{Jiang2020}, and 1.8/2.0~(EM7.5)~\cite{Arthuis2024} interactions. With these Hamiltonians, we obtain the following charge radii (listed in the same order as the three interactions): 3.538(32), 3.479(4), and 3.541(6)~fm for $^{52}$Ca and 3.563(32), 3.502(5), and 3.565(7)~fm for $^{54}$Ca (see also Ref.~\cite{Lellinger2026}).

The input amplitude for the DWIA calculation was computed as $\langle J'(A-1) | \hat{\psi}_{nlj}(r) | J(A) \rangle$ with $\hat{\psi}_{nlj}(r) = \psi_{nlj}(r)\hat{a}_{nlj}$, where $\psi_{nlj}(r)$ is the radial single-particle wave function for $nlj$ orbital with either HF or NAT basis~\cite{Tichai2019,Hoppe2021}. We choose to focus the discussion on the case of $^{54}$Ca$(p,pn)$$^{53}$Ca shown in Fig.~\ref{pmdEX_54Ca_np3_np1} and $^{54}$Ca$(p,2p)$$^{53}$K shown in Fig.~\ref{pmdEX_54Ca_pd3_ps1} (see Appendix for all reaction channels). From these examples we observe that the choice of interaction has little impact on the obtained momentum distributions; quantitatively, the mean relative difference (max--min/average) between the momentum distribution curves obtained with IMSRG and the three presented interactions are around 5.9\% for the neutron-knockout channels and around 11.3\% for the proton-knockout channels.
Additional checks were performed regarding the effects of the chosen basis, HF or NAT, and on the frequency ($\hbar\omega$) dependence. We find that the two different bases give similar momentum distributions. On the other hand, we observe a large frequency dependence. In the case of the proton knockout from $^{54}$Ca, i.e., deeper-bound nucleons, one finds that the frequency dependence is substantially reduced reflecting that both HF and NAT procedures optimize the deeply bound single-particle orbitals. A detailed analysis on the frequency dependence can be found in the Appendix.

To gain more insights, we examine an additional way to compute the single-particle radii. 
We calculate the single-particle radii starting from the one-body squared radius operator, where we turn on the matrix elements only for a given total angular momentum, orbital angular momentum, and isospin projection.
Applying the consistent IMSRG transformation of the operator and taking the ground-state expectation value, one can approximately compute the single-particle radii including the IMSRG-induced two-body contributions.
The reduced frequency dependence of the computed radii indicates that the notable frequency dependence (and the corresponding transition amplitude) may be partially attributed to missing IMSRG-induced one-particle-two-hole contributions.
Additionally, the use of consistent interactions in the structure and reaction calculations is expected to further reduce the frequency dependence.

\subsection{SCGF and DWIA}

The second set of momentum distributions were calculated using overlap functions from SCGF simulations using up to 14 major harmonic oscillator shells as described in Refs.~\cite{Cip2025prc,Soma2020}. We used the same three Hamiltonians employed in the VS-IMSRG calculations.

In SCGF theory, dynamical correlations are computed using the optimized reference states basis detailed in Refs.~\cite{Rocco2018,Raimondi2019,Bar2022prc}. These states have similar characteristics to natural orbitals for describing spatial distributions but are also tuned to encode information on nucleon separation energies. We perform our simulation using the so-called algebraic diagrammatic construction at third order, ADC(3) (see Ref.~\cite{Bar2017NLP,Raimondi2019}), which ensures convergence of nuclear radii with respect to many-body truncations. For the overlap functions representing the single-particle transition studies here, the ADC(3) includes full contributions from two-hole--one-particle (2h1p) and two-particle--one-hole (2p1h) states. Hence, the residual dependence on the oscillator frequency is only associated to model space convergence. To estimate the latter uncertainty we followed Ref.~\cite{Art2020prl} and performed simulations at different values of $\hbar\omega$ and model space sizes to bracket the converged model space limit, see the Appendix for details.

The resulting variation in the overlap functions was used to compute the SCGF bands, which give a conservative estimate of the model space truncation uncertainties. We estimated an error of at most 1.5\% on predicted nuclear radii~\cite{Soma2020,Bro2026tbp} for the three interactions considered.
Theoretical uncertainties associated with nuclear forces can be judged comparing results from different interactions. For example, the calculated charge radii of $^{52}$Ca are 3.580(19), 3.566(5), and 3.613(9)~fm, while the calculated charge radii of $^{54}$Ca are 3.616(12), 3.594(4), and 3.641(7)~fm, using the NNLO$_{\mathrm{sat}}$, $\Delta$N$^{2}$LO$_{\rm GO}$\,(394), and 1.8/2.0~(EM7.5) interactions, respectively; the difference of squared charge radii between $^{48}$Ca and $^{52}$Ca ranges between 0.36 and 0.40~fm$^2$ for the three different interactions used here, compared to the experimental result of 0.530(5)~fm$^2$~\cite{GarciaRuiz2016}.

The momentum distribution curves with DWIA and SCGF are plotted in Figs.~\ref{pmdEX_54Ca_np3_np1}  and ~\ref{pmdEX_54Ca_pd3_ps1}. There is a good agreement with the experimental data and there is little discrepancy between the three interactions; quantitatively, the mean relative difference (max--min/average) across the full momentum range between the momentum distribution curves obtained with SCGF and the three interactions are around 2.6\% for both the neutron-knockout and the proton-knockout channels. The equivalent $R_{\mathrm{s.p.}}$ using the SCGF overlap functions are also plotted in Figs.~\ref{RMS_neutrons}  and~\ref{RMS_protons} using open star markers. 
Note that besides overlap functions, the SCGF simulations also provide consistent information on scattering states, however, the state-of-the-art ADC(3) many-body truncation is insufficient at the scattering energies of this work and advanced sampling methods will be necessary~\cite{Brolli2025,Brolli2026}.

\subsection{Mean-field calculations}
\begin{figure}[t]
    \centering
    \includegraphics[width=0.495\textwidth]{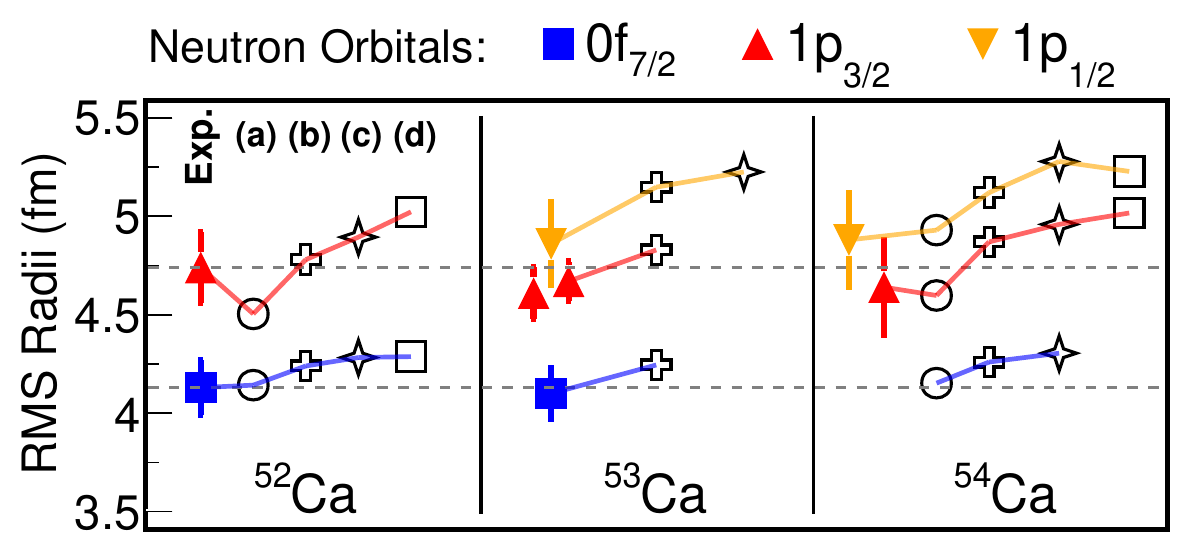}\\
    \caption{Rms radii of single-particle $0f_{7/2}$, $1p_{3/2}$, and $1p_{1/2}$ neutron orbitals in $^{52-54}$Ca. Experimental data for $^{52}$Ca come from Ref.~\cite{Enciu2022} and horizontal lines mark the $\nu 1p_{3/2}$ and $\nu 0f_{7/2}$ rms radii values of $^{52}$Ca for comparison with the present work. The effective single-particle radii extracted from experimental data (Exp.) are compared to the (a) HF (SKM$^{\ast}$), (b) RHF (PKA1), (c) SCGF [1.8/2.0~(EM7.5)], and (d) VS-IMSRG [1.8/2.0~(EM7.5)] predictions, see text for details. The data points and calculations are connected by lines to guide the eye.}
    \label{RMS_neutrons}
\end{figure}

\begin{figure*}[t]
    \centering
    \includegraphics[width=0.7\textwidth]{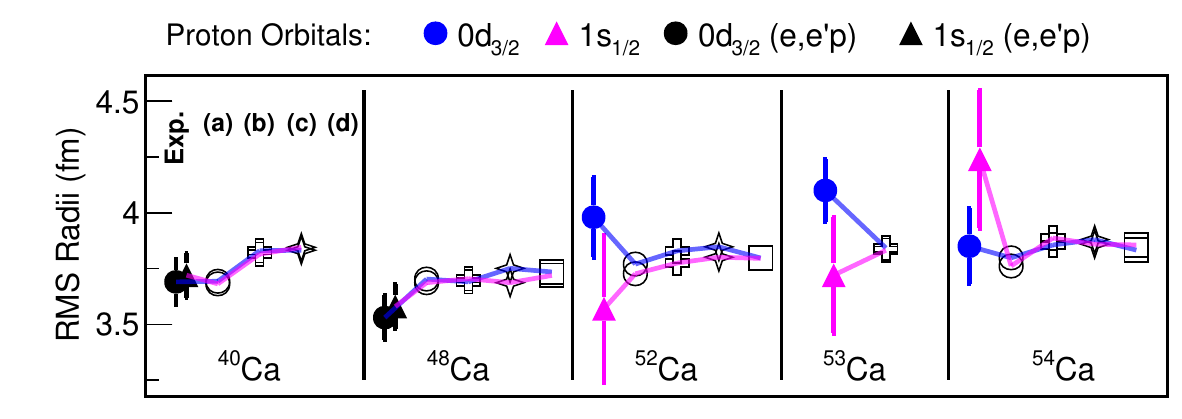}
    \caption{Rms radii of the 1$s_{1/2}$ and 0$d_{3/2}$ proton single-particle orbitals of $^{40,48,52,53,54}$Ca. Our extracted data points based on the momentum distributions are found in the last three columns for $^{52,53,54}$Ca (Exp., colored symbols). For $^{40,48}$Ca, the experimental data points (black symbols) come from (e,e'p) experiments~\cite{Kramer1989,Kramer2001}. The theoretical calculations are labeled as follows: (a) HF (SKM$^{\ast}$), (b) RHF (PKA1), (c) SCGF [1.8/2.0~(EM7.5)], and (d) VS-IMSRG [1.8/2.0~(EM7.5)] calculations, see text for details. The data points and calculations are connected by lines to guide the eye.}
    \label{RMS_protons}
\end{figure*}

The size of single-particle orbitals in neutron-rich Ca isotopes was also obtained from two mean-field calculations. These are compared to our data in Figs.~\ref{RMS_neutrons} and~\ref{RMS_protons}.
First, the Skyrme-Hartree-Fock (HF) calculations~\cite{Inakura2006,Horiuchi2012,Horiuchi2020} for $^{40,48,50,52}$Ca with the SkM$^\ast$ interaction~\cite{Bartel1982}. Pairing correlations are not taken into account in the HF calculations, but their effect on nuclear radii is small for a spherical ground state~\cite{Horiuchi2016}. To preserve the spherical symmetry, we employ the equal-filling approximation~\cite{Beiner1975}, i.e., when the Fermi level with angular momentum $j$ is occupied partially by $m$ nucleons with $m<2 j+1$, we assign an equal occupation probability to these $m$ nucleons as $m/(2 j + 1)$. 

In the second case, relativistic Hartree Fock (RHF) calculations were performed using the PKA1 interaction~\cite{Liu2020,Long2007PKA1}. This interaction reproduces the subshell closures at $N=32,34$, the separation energies for Ca isotopes~\cite{Li2016, Liu2020}, and the charge radii of $^{48}$Ca and $^{52}$Ca: 3.492~fm and 3.546~fm, respectively, leading to a difference of $\delta\braket{r^2}_{ch}^{48,52}$ = 0.38~fm$^2$. The results of these calculations for the size of the single-particle orbitals are found in good agreement with the values extracted from the experimental momentum distributions, as shown in Figs.~\ref{RMS_neutrons} and~\ref{RMS_protons}.


\section{Discussion}
For the neutrons, a remarkable feature is that the valence $\nu 1p_{3/2}$ and $\nu 1p_{1/2}$ orbitals for $^{53,54}\text{Ca}$ are significantly larger than the deeply bound $\nu 0f_{7/2}$ core orbital by approximately $0.48-0.78\text{ fm}$. This substantial spatial extension is consistent with previous findings for $^{52}\text{Ca}$~\cite{Enciu2022} and the prediction from Ref.~\cite{Bonnard2016}. For the $\nu1p$ orbitals, the valence $\nu 1p_{1/2}$ orbitals systematically extend slightly further than their $\nu 1p_{3/2}$ partners, owing to the progressive weakening of nucleon binding as neutrons are added. 
Our {\it ab initio} VS-IMSRG and SCGF computations as well as the relativistic Hartree-Fock results reproduce this difference in rms radii, 0$f_{7/2}$--1$p$ and 1$p_{3/2}$--1$p_{1/2}$, in very good agreement with the experimentally extracted values within 1$\sigma$, while the Hartree-Fock calculations for $^{52}\text{Ca}$ agree with our experimentally extracted values within 2$\sigma$.

For the protons, the extracted size of the $\pi 0d_{3/2}$ orbital remains constant across the $^{52-54}\text{Ca}$ isotopes, with a weighted mean value of 4.00(9)~fm. However, the valence $\pi 1s_{1/2}$ orbital experiences a marginal increase in its rms radius moving toward $^{54}\text{Ca}$, with a weighted mean value of 3.87(17)~fm. Both proton valence orbitals, $\pi 1s_{1/2}$ and $\pi 0d_{3/2}$, reveal an increase in size in $^{52,53,54}\text{Ca}$ when compared to the stable $^{40}\text{Ca}$ and $^{48}\text{Ca}$ measured in past $(e,e'p)$ experiments~\cite{Kramer1989,Kramer2001}. The $(e,e'p)$ studies resulted in 3.69(10) and 3.53(10)~fm for the $\pi 0d_{3/2}$ and 3.72(10) and 3.58(10)~fm for the $\pi 1s_{1/2}$ orbitals of $^{40}$Ca and $^{48}$Ca, respectively; the $(e,e'p)$ data points are also plotted for comparison in Fig.~\ref{RMS_protons}. These result in a mean increase of 0.47(13)~fm for $0d_{3/2}$ and 0.29(20)~fm for $1s_{1/2}$ along the $^{52-54}$Ca relative to the $^{48}$Ca core. From the global systematics of the theoretical models, we observe a slight shrinking of the $sd$-shell proton single-particle orbitals from $^{40}\text{Ca}$ to $^{48}\text{Ca}$ followed by a steady and continuous expansion from $^{48}\text{Ca}$ up to $^{54}\text{Ca}$. It is worth noting that the calculated increase in the $0d_{3/2}$ and $1s_{1/2}$ proton orbital size from $^{48}$Ca to $^{52-54}$Ca is smaller than the increase observed in the extracted values from the momentum distributions. Only values of 0.115(7)~fm (SCGF) and 0.08~fm (VS-IMSRG) are found for the increase of the $0d_{3/2}$ rms radius and 0.146(10)~fm (SCGF) and 0.11~fm (VS-IMSRG) are found for the increase of the $1s_{1/2}$ rms radius.

A systematic structural ``swelling'' provides a mechanism for the steep, non-linear increase in bulk charge radii observed experimentally after the $N=28$ shell closure---a phenomenon that has challenged traditional microscopic calculations~\cite{Horiuchi2020,GarciaRuiz2016,Heinz2024}. As the spatially extended $\nu 1p$ neutron orbitals are filled beyond $N=28$, their peripheral wave functions trigger a core swelling effect that pushes the underlying proton distribution outward according to Ref.~\cite{Bonnard2016}. While this study supports the large 0$f_{7/2}$--1$p$ rms radii difference in the neutron orbitals as predicted by Ref.~\cite{Bonnard2016}, the evolution of the proton orbitals from $^{48}$Ca to $^{52-54}$Ca remains inconclusive. The statistical uncertainties associated with our extracted $\pi 1s_{1/2}$ and $\pi 0d_{3/2}$ rms radii remain relatively large, which is primarily attributed to the lower cross sections and statistics inherent to the $(p,2p)$ reaction channels. Moreover, for a fair comparison of the size of single-particle orbitals from $^{48}$Ca up to $^{52-54}$Ca, the same experimental probe should be used instead of the different ones, $(e,e'p)$ and $(p,2p)$, in the present work.

Within the DWIA, the single-particle cross sections are also highly sensitive to the radial parameter used for the wave function of the knocked-out nucleon. By optimizing the radial parameter to reproduce the measured momentum distributions in this study, we directly impact the single-particle cross sections and the resulting theoretical exclusive and inclusive cross sections. With the quenching factor defined as $\sigma^{\mathrm{exp}}$/$\sigma^{\mathrm{th}}$ for the inclusive cross sections, where $\sigma^{\mathrm{th}}_{\mathrm{incl.}}$ = $\sum SF\cdot\sigma_{\mathrm{sp}}$, summed over all populated bound states, we find no quenching factors, 0.96 and 1.0 for neutron knockout from $^{53}$Ca and $^{54}$Ca, respectively. The spectroscopic factors were calculated using shell model calculations with the $pf$-shell part of the PFSDG-U interaction assuming a $^{40}$Ca core~\cite{Nowacki2016} for $^{53}$Ca and with the GXPF1Bs interaction reported in Ref.~\cite{Chen2019} for $^{54}$Ca. For the proton knockout channels, the quenching factor for the inclusive cross section for $^{52}$Ca cannot be calculated due to the unresolved final states populated in $^{51}$K, while for $^{53}$Ca and $^{54}$Ca quenching factors of 0.65 and $0.51-0.66$ were obtained, respectively, using the spectroscopic factors reported in Refs.~\cite{Sun,Enciu2024}. As a result, we obtain an isospin dependence of the quenching factors, in agreement with other findings in the literature, see Refs.~\cite{Gade2014,Holl2019,Gade2021,Aumann2021,Gomez-Ramos2023}.

\section{CONCLUSIONS}
Exclusive momentum distributions were measured in this work via quasi-free $(p,2p)$ and $(p,pn)$ knockout reactions from $^{52-54}$Ca at $\sim$230~MeV/nucleon. Based on the DWIA formalism, we show sensitivity of the exclusive momentum distributions to the radial extension of the single-particle orbitals corresponding to the knocked-out nucleon, and the size of single-particle $\nu 0f_{7/2}$, $\nu 1p$, $\pi 0d_{3/2}$, and $\pi1s_{1/2}$ orbitals in neutron-rich Ca isotopes are extracted for the first time. To this end, we have combined microscopic {\it ab initio} calculations with DWIA calculations. We found a difference between the rms radius of $0f_{7/2}$ and $1p$ neutron orbitals ranging between 0.48 and 0.78~fm, persisting throughout $^{52}$Ca to $^{54}$Ca. This size difference is well reproduced by {\it ab initio} VS-IMSRG and SCGF calculations. For the $sd$-shell proton orbitals we obtain a mean increase of 0.47(13)~fm and 0.29(20)~fm for the $0d_{3/2}$ and $1s_{1/2}$ orbitals, respectively, in $^{52-54}$Ca relative to the $^{48}$Ca values obtained from $(e,e'p)$ measurements. Although such large differences in the $0d_{3/2}$ and $1s_{1/2}$ are not reproduced by our theoretical calculations, a 2$\sigma$ agreement holds. Given the large statistical uncertainties, this study calls for further systematic experiments with improved statistics and missing-momentum resolution for the Ca isotopes between $A=40-54$ using proton-induced nucleon reactions, $(p,pN)$, as well as benchmarking $(p,2p)$ against $(e,e'p)$ on the stable isotopes.

Looking forward, reducing the statistical and systematic uncertainties inherent to weakly populated proton channels will require next-generation experimental tracking setups with enhanced momentum resolution. From a theoretical standpoint, the next major step lies in moving away from standard, localized Woods-Saxon potential approximations. This will require combining {\it ab initio} transition amplitudes directly into the reaction theory framework for a coherent method of probing nuclear states.  Moreover, {\it ab initio} computations of the nuclear self-energy may inform on the optical potentials to be used in DWIA simulations~\cite{Waldecker2011,Idini2019} and help mitigate inconsistencies between the structure and reactions parts of this analysis. Systematically expanding this technique along isotopic chains will help develop the momentum distributions after quasi-free knockout reactions from a qualitative probe to a quantitative one, refining our understanding of neutron and matter radii of atomic nuclei. \\

\begin{acknowledgments}
We are grateful for the support of the RIKEN Nishina Center accelerator staff in the delivery of the primary beam and the BigRIPS team for preparing the secondary beams. The development of MINOS has been supported by the European Research Council through the ERC Grant No. MINOS-258567. M. E., A. O., A. S., T. A., I. G., C. L., D. R., H. T., V. W., and L. Z. acknowledge the support from the Deutsche Forschungsgemeinschaft (DFG, German Research Foundation) -- Project-ID 279384907 -- SFB 1245. K. O. acknowledges the support by Grants-in-Aid for Scientific Research from the JSPS (No.~JP21H00125). K. Y. acknowledges the support by Grants-in-Aid for Scientific Research from the JSPS (No.~JP20K14475 and No.~JP25K17400). M. H., T. M., and A.S.~were supported in part by the European Research Council (ERC) under the European Union's Horizon 2020 research and innovation programme (Grant Agreement No.~101020842). T. M. is supported by JST ERATO Grant No.~JPMJER2304, Japan. A. P. is funded by Grant CEX2020-001007-S funded by MCIN/AEI/10.13039/501100011033 and PID2021-127890NB-I00. M. H. is supported by the Laboratory Directed Research and Development Program of Oak Ridge National Laboratory, managed by UT-Battelle, LLC, for the U.S.\ Department of Energy. F. B. was supported by the RIKEN Special Postdoctoral Researcher Program. M. E. was supported by an ETH Z\"urich Postdoctoral Fellowship during manuscript writing and discussions reaching the final conclusions of the presented study. D. S. acknowledges the support of GINOP-2.3.3-15-2016-00034. S. B. and C. B. acknowledge the use of the DiRAC Data Intensive service (DIaL3) at the University of Leicester, managed by the University of Leicester Research Computing Service on behalf of the STFC DiRAC HPC Facility (www.dirac.ac.uk). The DiRAC service at Leicester was funded by BEIS, UKRI and STFC capital funding and STFC operations grants. DiRAC is part of the UKRI Digital Research Infrastructure. This research used resources of the Oak Ridge Leadership Computing Facility located at Oak Ridge National Laboratory, which is supported by the Office of Science of the Department of Energy under contract No.~DE-AC05-00OR22725. The authors gratefully acknowledge the Gauss Centre for Supercomputing e.V.\ (www.gauss-centre.eu) for providing computing time through the John von Neumann Institute for Computing (NIC) on the GCS Supercomputer JUWELS at Jülich Supercomputing Centre (JSC). For the IMSRG calculations, the \texttt{NuHamil}~\cite{NuHamil} and \texttt{imsrg++}~\cite{imsrg++} codes were used to generate the necessary Hamiltonian matrix elements and to perform the VS-IMSRG evolution, respectively.
\end{acknowledgments}

\raggedbottom

\appendix
\section{Details on IMSRG calculations}
In this work, the squared single-particle radius is computed as 
\begin{equation}
R^{2}_{\mathrm{s.p.}, \, \mathrm{IMSRG}} = \int^{\infty}_{0} dr \, r^{4} |\langle J'(A-1) | \hat{\psi}_{nlj}(r) | J(A) \rangle|^{2},
\end{equation}
with $\int^{\infty}_{0} dr \, r^{2} |\langle J'(A-1) | \hat{\psi}_{nlj}(r) | J(A) \rangle|^{2}$ normalized to 1.
Using the 1.8/2.0 (EM)~\cite{Hebeler2011} interaction, we observe a significant frequency dependence of $\langle J'(A-1) | \hat{\psi}_{nlj}(r) | J(A) \rangle$ both with HF and NAT orbitals. The basis frequency is an artificial parameter, and any residual dependence on it is usually related to many-body and model space truncations. 
As discussed in the main text, in this work, we select $\hbar\omega = 12$~MeV, as the computed point-proton and point-neutron radii show a better convergence. 

To guide future development to reduce the frequency dependence, we examine two other ways to compute the single-particle radii. The first one is computed as the diagonal one-body matrix element of the consistently evolved point-proton or point-neutron radius operator.
It is found that $R^{2}_{s.p., \, \mathrm{IMSRG}}$ and single-particle radii computed in this way are very similar and show a similar frequency dependence.
In the second definition, we consider complementary computations of single-particle radii starting from the one-body squared radius operator, where we turn on the matrix elements only for a given total angular momentum, orbital angular momentum, and isospin projection.
Taking the ground-state expectation value of the operator, one can approximately compute the single-particle radius including the IMSRG-induced two-body components. From this approach, we find that inclusion of the IMSRG-induced two-body term is essential to mitigate the frequency dependence.
The current definition of $\langle J'(A-1) | \hat{\psi}_{nlj}(r) | J(A) \rangle$ involves only $\langle J'(A-1)| \hat{a}_{nlj} | J(A) \rangle$.
However, the IMSRG transformation of the annihilation operator will induce many-body terms:
\begin{equation}
U\hat{a}_{nlj}U^{\dag} = \hat{a}_{nlj} + \sum_{n'l'j'} \sum_{n''l''j''} \hat{a}^{\dag}_{n''l''j''} \hat{a}_{n'l'j'} \hat{a}_{nlj} + \ldots.
\end{equation}
This suggests that inclusion of the first induced term is significant to reduce the frequency dependence and that further development will be needed.

\section{Details on SCGF calculations}

\begin{figure*}[t]
    \centering
    \includegraphics[width=0.8\textwidth]{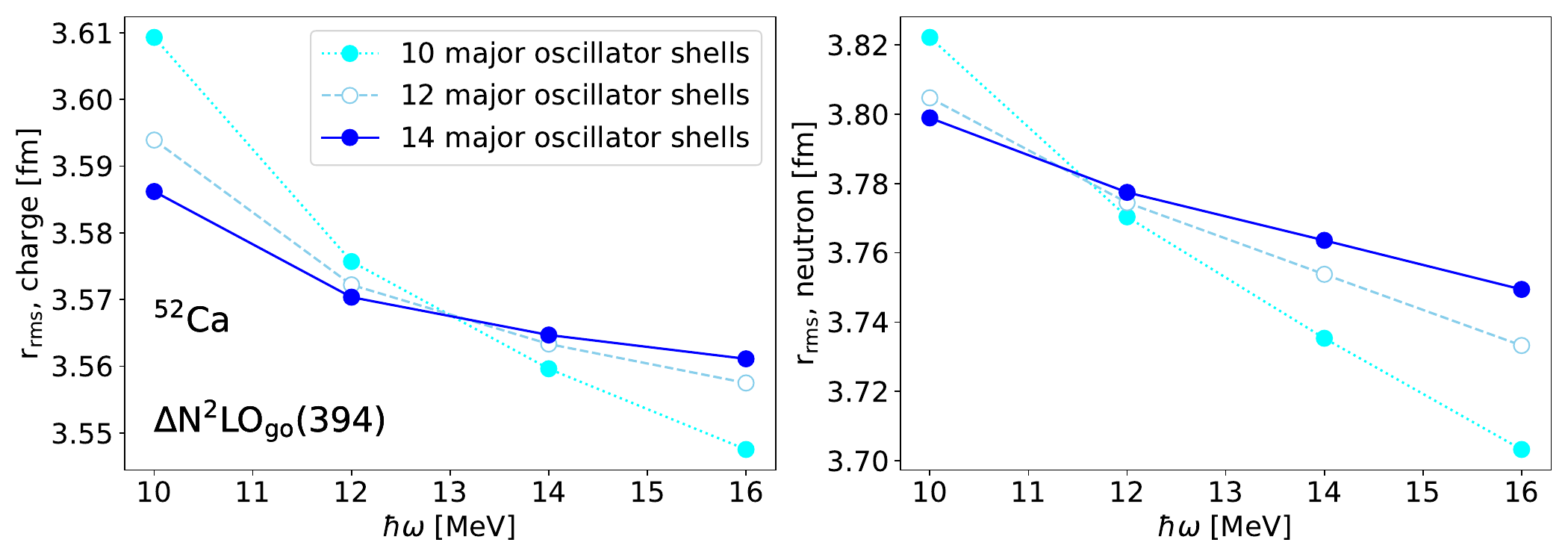}
    \caption{Dependence of the computed charge (left) and neutron (right) rms radii for $^{52}$Ca on the oscillator frequency and the number of major oscillator shells. For the purpose of estimating uncertainties, upper and lower frequencies of $\hbar\omega=$~12 and 16~MeV were used for protons and $\hbar\omega=$~10 and 14~MeV for neutrons. The $\Delta$N$^2$LO$_{\mathrm{GO}}$\,(394) interaction~\cite{Jiang2020} was used here.}
    \label{Fig:hw_dependence}
\end{figure*}
The one-body Green's function (or propagator) associated to the ground state of an $A$-nucleon system takes the following form in spherical coordinates~\cite{Bar2017NLP}:
\begin{equation}
\begin{split}
&g_{lj\tau}(r,r';E)\\
 &= \sum_k
   \frac{ \langle\Psi^A_{\rm g.s.}|\hat\psi^\dagger_{lj\tau}(r)|\Psi^{A-1}_k\rangle 
          \langle\Psi^{A-1}_k|\hat\psi_{lj\tau}(r')|\Psi^A_{\rm g.s.}\rangle}
        {E + (E^{A-1}_k - E^A_{\rm g.s.}) - i\eta}\\
 &\quad + \sum_n
   \frac{ \langle\Psi^A_{\rm g.s.}|\hat\psi_{lj\tau}(r')|\Psi^{A+1}_n\rangle 
          \langle\Psi^{A+1}_n|\hat\psi^\dagger_{lj\tau}(r)|\Psi^A_{\rm g.s.}\rangle}
        {E - (E^{A+1}_n - E^A_{\rm g.s.}) + i\eta}\,.
\end{split}
\label{eq:gprop}
\end{equation}
where $l$, $j$ and $\tau$ are the orbital and total angular momentum, and isospin projection quantum numbers of the transferred nucleon and the sums over $k$~($n$) runs over the discrete and continuum spectrum of the particle removed (particle attached) system. The overlap functions describing nucleon knockout from specific single-particle states can be read directly from the residues of Eq.~(\ref{eq:gprop}). 
We computed the propagator for $^{52}$Ca and $^{54}$Ca and extracted the proton and neutron removal functions from the first term on the right-hand side of Eq.~(\ref{eq:gprop}). The transition wave functions from $^{53}$Ca and $^{55}$Sc to the ground states of $^{52}$Ca and $^{54}$Ca were available from the second term.  Note that for nucleon attachment to an $A+1$ state in the continuum, the residues of the second term on the right-hand side of Eq.~(\ref{eq:gprop}) also provide elastic scattering wave functions.

\begin{figure*}[ht]
  \centering
    \begin{minipage}[t]{0.48\textwidth}
    \centering
    \includegraphics[width=\linewidth]{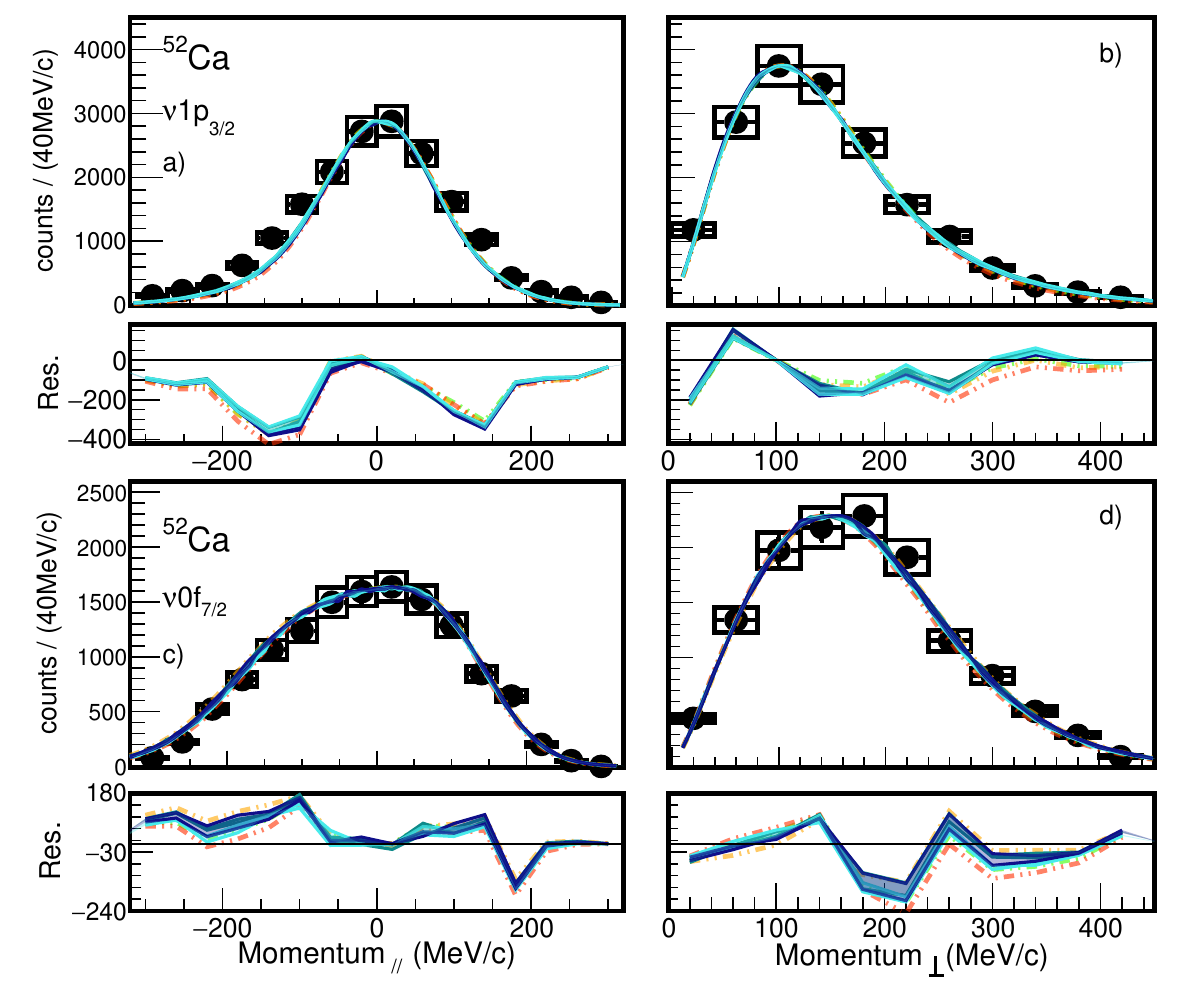}
    \caption{Plots for the $^{52}$Ca$(p,pn)$$^{51}$Ca reaction channel. Legend same as Fig.~\ref{pmdEX_54Ca_np3_np1}.}
    \label{pmd_52Ca_ppn_residuals}
  \end{minipage}\hfill
  \begin{minipage}[t]{0.48\textwidth}
    \centering
    \includegraphics[width=\linewidth]{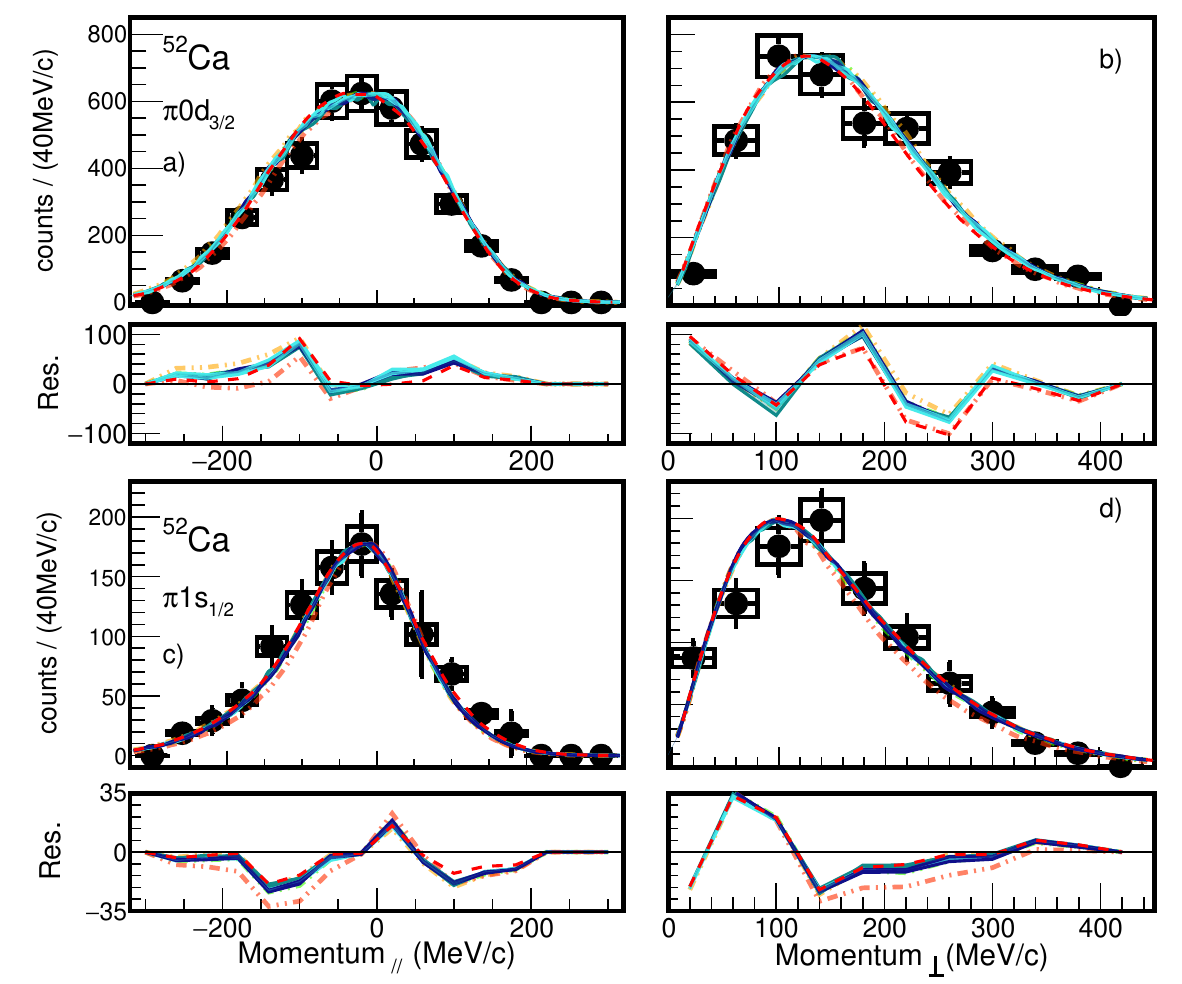}
    \caption{Plots for the $^{52}$Ca$(p,2p)$$^{51}$K reaction channel. Legend same as Fig.~\ref{pmdEX_54Ca_np3_np1}.}
    \label{pmd_52Ca_p2p_residuals}
  \end{minipage}
\end{figure*}

In SCGF theory, the propagator~(\ref{eq:gprop}) is obtained by solving the Dyson equation, schematically  $g(E) = g^0(E) + g^0(E) \, \Sigma^\star(E)\, g(E)$, where $g^0(r,r';E)$ is the Green's function for a freely propagating nucleon, subject only to kinetic energy.  The nuclear correlations and dynamics are contained in the self-energy operator $\Sigma^\star(r,r';E)$ which is a bound object and can be efficiently expanded in a harmonic oscillator basis. We compute $\Sigma^\star(r,r';E)$ in a space of 14 major shells, however, the Dyson equation was solved without truncating the single-particle model space (up to 150 oscillator shells were used to reach convergence) and including complete kinetic energy terms. This guarantees the proper asymptotic behavior, with exponential tails, for the computed overlap functions.
The ADC(2) many-body truncation for the self-energy includes non-interacting 2h1p and 2p1h contributions and it was found to be already sufficient to provide precise rms radii~\cite{Roc2019,Soma2020}. However, we performed our simulation using the more complete ADC(3) scheme. Hence, residual dependence on the oscillator frequency $\hbar\omega$ reflects the model space truncations in the computation of $\Sigma^\star(r,r';E)$. The typical dependence on the model space size and frequency is displayed in Fig.~\ref{Fig:hw_dependence} for the case of charge and neutron radii on $^{52}$Ca: For small values of $\hbar\omega$ (corresponding to a large oscillator length, $b^2= \hbar/ (m \omega)$) computed radii converge in a monotonically decreasing fashion when increasing the number of major shells. The trend becomes monotonically increasing for large values of $\hbar\omega$ so that curves cross near the converged values for the complete model space. Similarly to Ref.~\cite{Art2020prl}, we studied the convergence of neutron radii as a function of the harmonic oscillator frequency for each isotope. We chose the optimal value of $\hbar\omega$ as the one for which computed radii are least sensitive to the model space truncation. Optimal oscillator frequencies were found to be in the range $\hbar\omega=14-16$~MeV for the NNLO$_{\mathrm{sat}}$ interaction~\cite{Ekstrom2015}, depending on the isospin, and within $\hbar\omega=12-14$~MeV for the $\Delta$N$^2$LO$_{\mathrm{GO}}$\,(394)~\cite{Jiang2020} and 1.8/2.0 (EM7.5)~\cite{Arthuis2024} interactions. We then computed the overlaps and the corresponding DWIA momentum distributions for the largest model space (14 major shells) but varying $\hbar\omega$ within approximately $\pm 2~\mathrm{MeV}$ around this optimal value. Given the large range on $\hbar\omega$, the results of this variations can be taken as a conservative estimate of the theoretical error associated with the model space truncation. These uncertainties are shown as bands in Figs.~3 and~4 of the main text and in Figs.~\ref{pmd_52Ca_ppn_residuals}  to~\ref{pmd_52Ca_p2p_residuals} here.

\section{Complete plots for comparison of calculated momentum distributions with experimental data}

The complete set of plots for the comparison of experimental exclusive momentum distributions and DWIA calculations using {\it ab-initio}-based transition amplitudes is shown in this section. The plot format and legend are the same as in Fig.~\ref{pmdEX_54Ca_np3_np1}.
Figure~\ref{pmd_52Ca_ppn_residuals} shows the exclusive momentum distribution plots for the $^{52}$Ca$(p,pn)$$^{51}$Ca reaction channel with the 1$p_{3/2}$ (panels \emph{a} and \emph{b}) and 0$f_{7/2}$ (panels \emph{c} and \emph{d}) neutron knock-out. Figure~\ref{pmd_52Ca_p2p_residuals} shows the exclusive momentum distribution plots for the $^{52}$Ca$(p,2p)$$^{51}$K reaction channel with the 0$d_{3/2}$ (panels \emph{a} and \emph{b}) and 1$s_{1/2}$ (panels \emph{c} and \emph{d}) proton knock-out. In each figure, the experimental data are shown with black filled circles, error bars for statistical uncertainties, and error boxes for systematic uncertainties. The momentum distributions are divided into the parallel (left panels) and perpendicular (right panels) projections. The theoretical calculations for the momentum distributions from DWIA together with the Woods-Saxon wave functions, SCGF (bands spanning upper-lower limits), and VS-IMSRG {\it ab initio} methods are also superimposed on the experimental data and plotted with different colors and line markers (see main text for details).
The SCGF and VS-IMSRG computations use three different interactions from chiral effective field theory~\cite{Ekstrom2015,Jiang2020,Arthuis2024}. The legend of each plot is the same as shown in Fig.~\ref{pmdEX_54Ca_np3_np1}. Each momentum distribution plot is accompanied by a residuals (Res.) plot below, showing the absolute difference between the theoretical curves and the experimental data. Residuals are given in the unit of counts per bin (bins of 40~MeV/c).

\section{Complete plots for the optimum radial parameter extraction}
\begin{figure*}[p]
\includegraphics[width=1\textwidth]{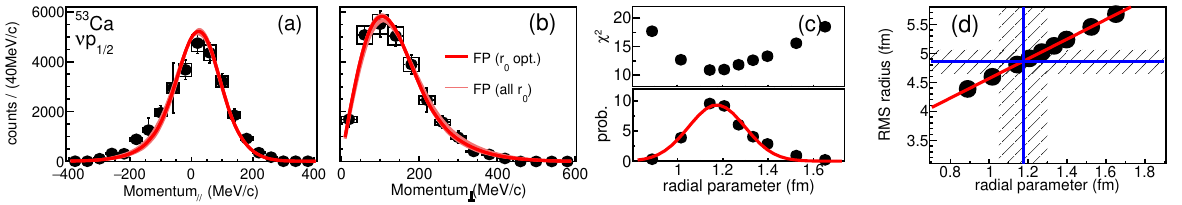}
    \includegraphics[width=1\textwidth]{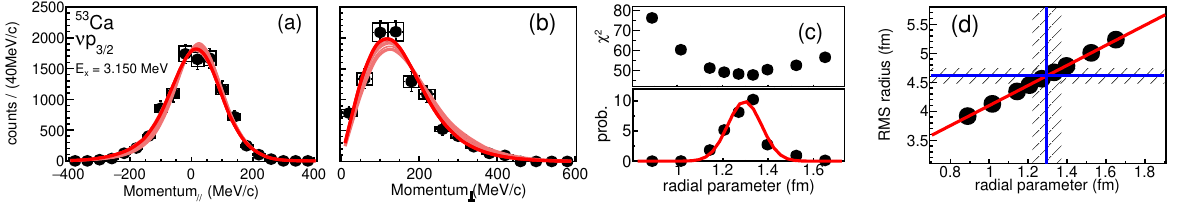}
    \includegraphics[width=1\textwidth]{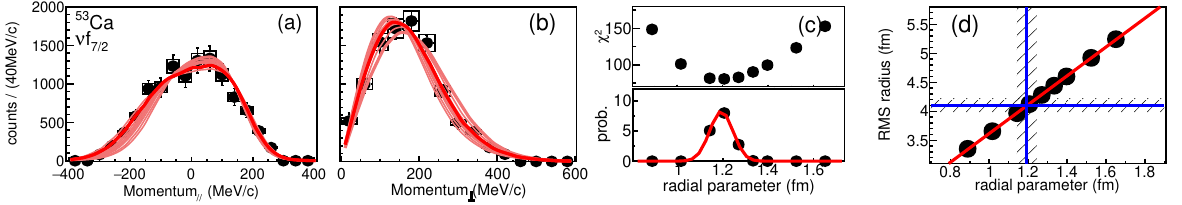}
    \caption{Plots for the $^{53}$Ca neutron knockout case including the $p_{1/2}$, $p_{3/2}$ (for final excitation energy of 3.150~MeV of $^{52}$Ca), and the $f_{7/2}$ valence neutron orbitals.}
    \label{pmd_53Ca_np3_np1_nf7}
    \centering
    \includegraphics[width=1\textwidth]{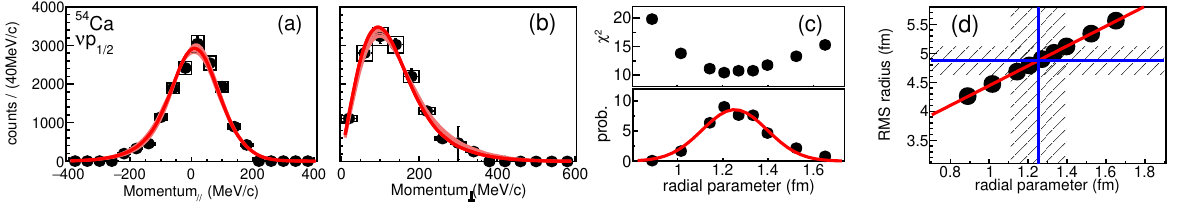}
    \includegraphics[width=1\textwidth]{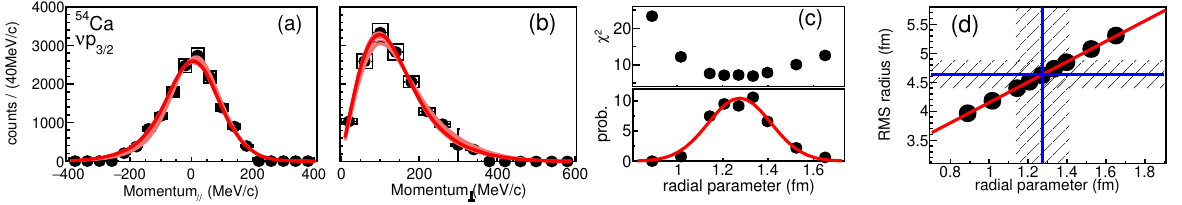}
    \caption{Plots for the $^{54}$Ca neutron knockout case including the $p_{1/2}$ and $p_{3/2}$ valence neutron orbitals.}
    \label{pmd_54Ca_np3_np1}
    \centering
    \includegraphics[width=1\textwidth]{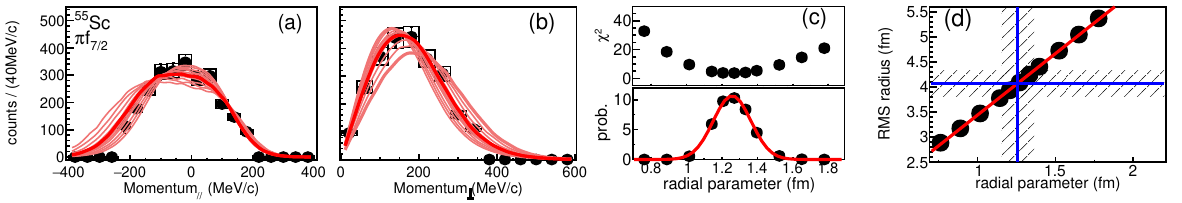}
    \caption{Plots for the $^{55}$Sc proton knockout case including the $f_{7/2}$ valence proton orbital.}
    \label{pmd_55Sc_pf7}
\end{figure*}

\begin{figure*}[p]
\centering
    \centering
    \includegraphics[width=1\textwidth]{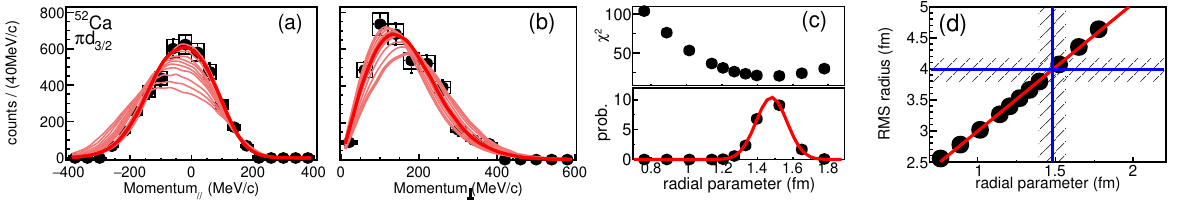}
    \includegraphics[width=1\textwidth]{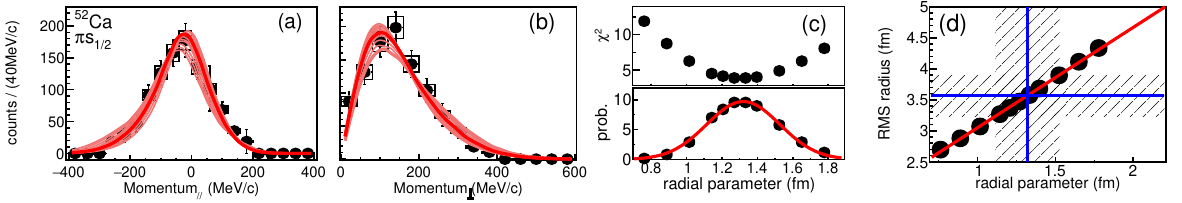}
    \caption{Plots for the $^{52}$Ca proton knockout case including the $s_{1/2}$, $d_{3/2}$valence proton orbitals.}
    \label{pmd_52Ca_pd3_ps1}
    \centering
    \includegraphics[width=1\textwidth]{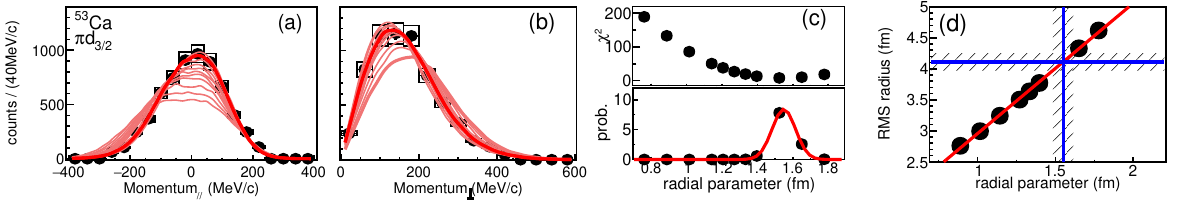}
    \includegraphics[width=1\textwidth]{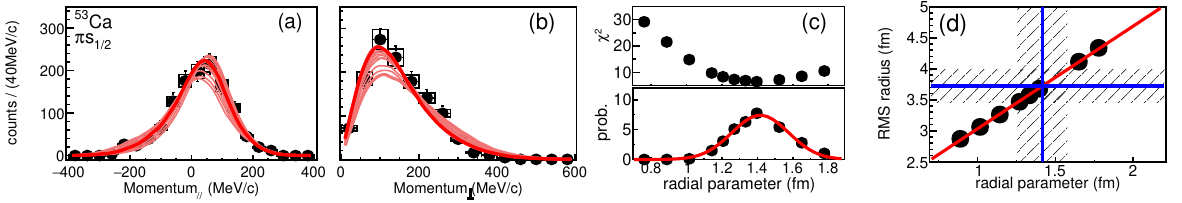}
    \caption{Plots for the $^{53}$Ca proton knockout case including the $s_{1/2}$, $d_{3/2}$valence proton orbitals.}
    \label{pmd_53Ca_pd3_ps1}
    \centering
    \includegraphics[width=1\textwidth]{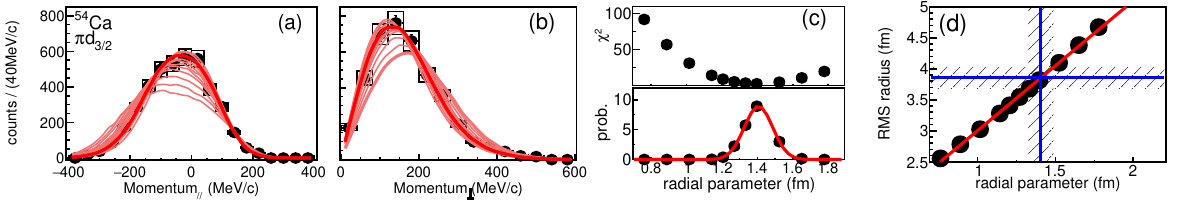}
    \includegraphics[width=1\textwidth]{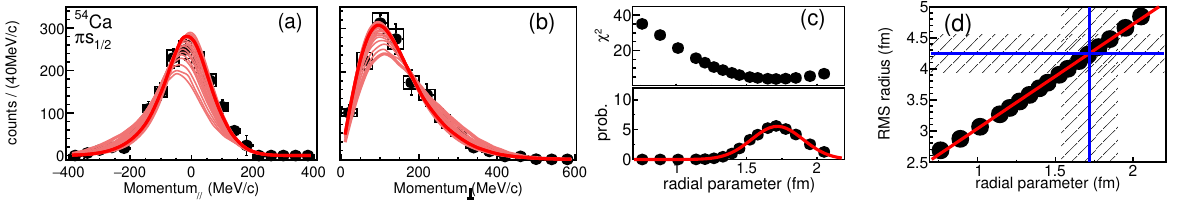}
    \caption{Plots for the $^{54}$Ca proton knockout case including the $s_{1/2}$, $d_{3/2}$valence proton orbitals.}
    \label{pmd_54Ca_pd3_ps1}  
\end{figure*}

Figures~\ref{pmd_53Ca_np3_np1_nf7} to~\ref{pmd_54Ca_pd3_ps1} show the complete set of plots used for the extraction of $R_{\mathrm{s.p.}}$. For each valence neutron and proton orbital, panel (a) shows the parallel momentum distribution and panel (b) shows the perpendicular momentum distribution. The experimental data is plotted with black filled circles, the statistical uncertainties are shown with vertical error bars, and the systematic uncertainties are shown with the boxes, the width of the box corresponding to the width of the chosen binning. With red curves, the DWIA calculations are shown for several tested radial parameters (see main text for details). In each case, one curve is shown with thicker red line -- this curve corresponds to the optimum value for the radial parameter. The $\chi^2$ (with both parallel and perpendicular momentum contributions) and the probability (prob.), defined as \hbox{probability $= K\cdot \exp(-\chi^2/2)$,} obtained after fitting each theoretical momentum distribution curve to the experimental data, are shown in panel (c). Panel (d) shows the relation between $R_{\mathrm{s.p.}}$ and the radial parameter for each orbital. The vertical blue line and the corresponding hatched area show the optimum radial parameter value, mean and sigma, respectively. The horizontal blue line and the corresponding hatched region show the corresponding $R_{\mathrm{s.p.}}$ mean and sigma extracted based on the linear correlation between $R_{\mathrm{s.p.}}$ and the radial parameter (red line).

\begin{figure*}[p]
    \centering
        \centering
        \includegraphics[width=0.49\linewidth]{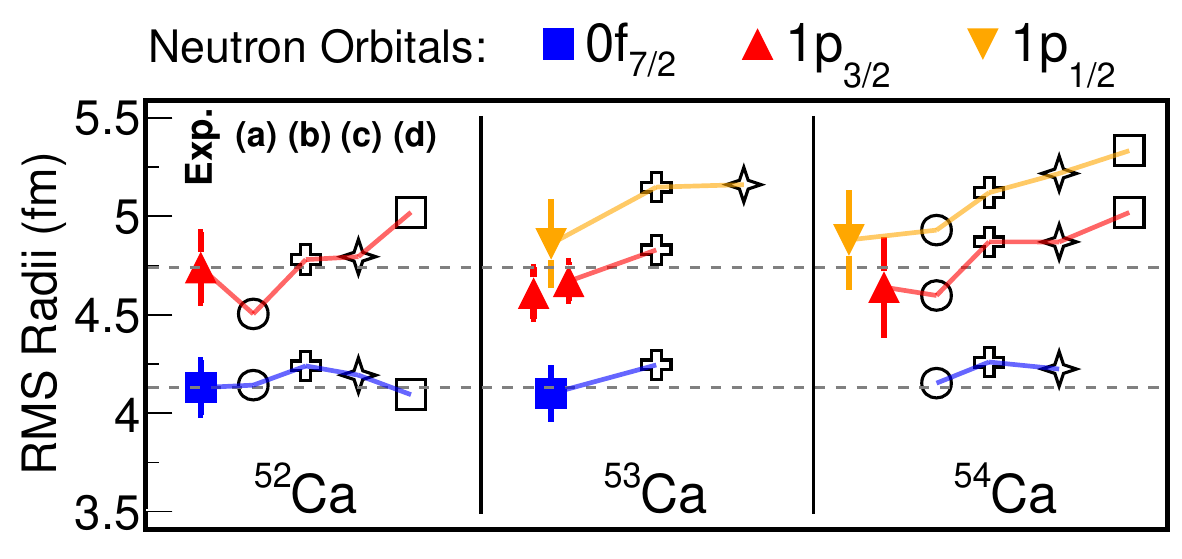}
        \caption{Same as Fig.~\ref{RMS_neutrons}, but (c) and (d) calculations use the $\Delta$N$^{2}$LO$_{\rm GO}$\,(394) interaction~\cite{Jiang2020}.}
        \label{fig:Neutron_Radii_sp_DNNLOGO394_plt}
        \centering
        \includegraphics[width=0.49\linewidth]{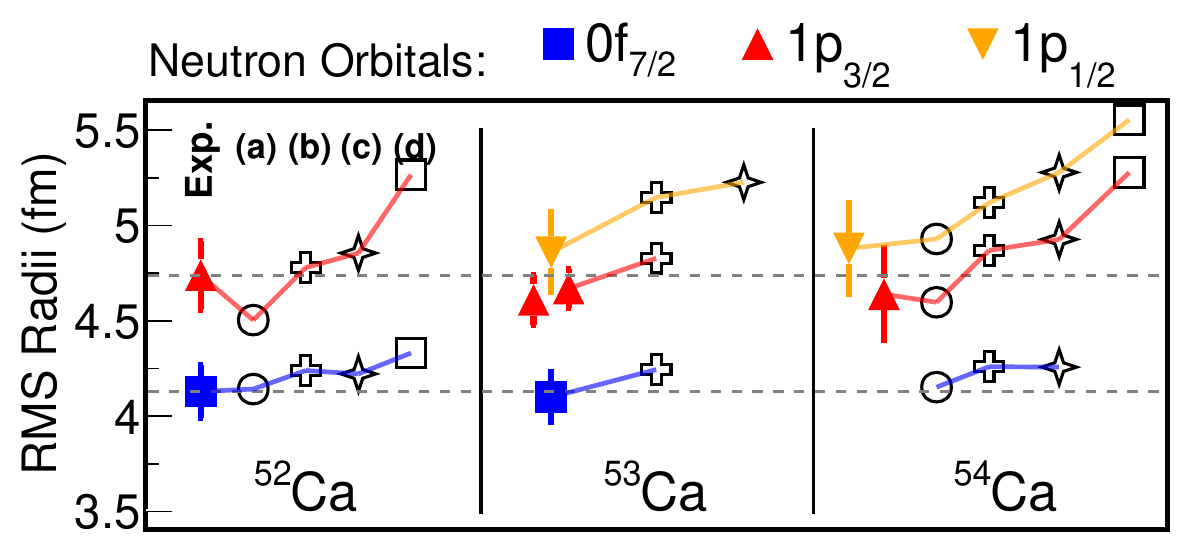}
        \caption{Same as Fig.~\ref{RMS_neutrons}, but (c) and (d) calculations use the NNLO$_{\mathrm{sat}}$ interaction~\cite{Ekstrom2015}.}
        \label{fig:Neutron_Radii_sp_NNLOsat_plt}
        \centering
        \includegraphics[width=0.7\linewidth]{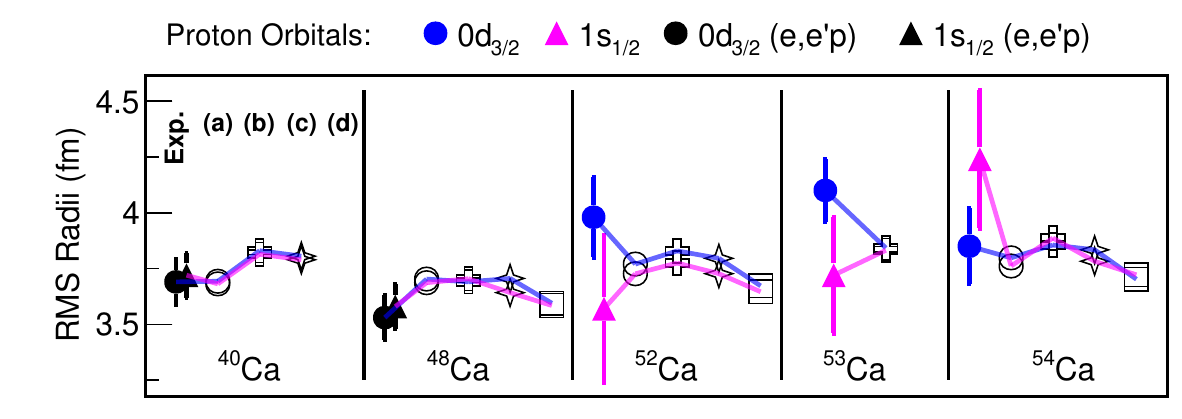}
        \caption{Same as Fig.~\ref{RMS_protons}, but (c) and (d) calculations use the $\Delta$N$^{2}$LO$_{\rm GO}$\,(394) interaction~\cite{Jiang2020}.}
        \label{fig:Proton_Radii_sp_DNNLOGO394_plt}
        \centering
        \includegraphics[width=0.7\linewidth]{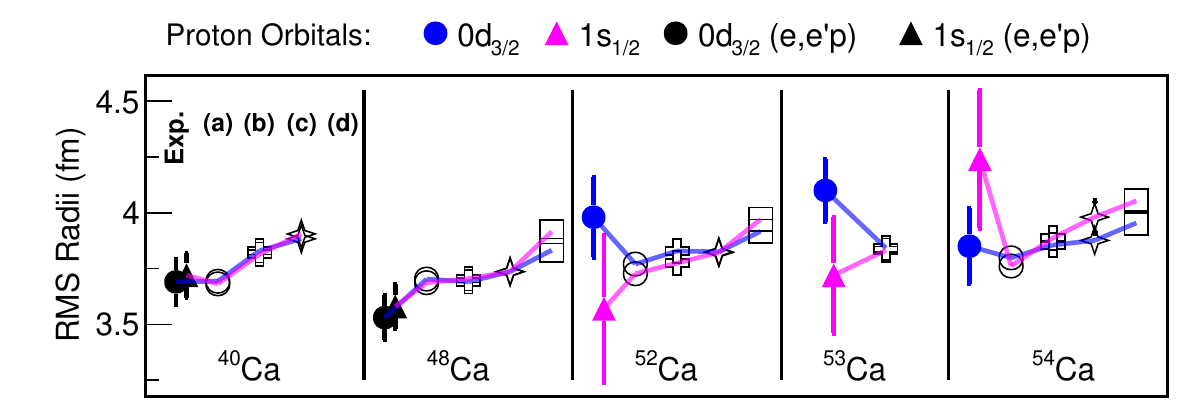}
        \caption{Same as Fig.~\ref{RMS_protons}, but (c) and (d) calculations use the NNLO$_{\mathrm{sat}}$ interaction~\cite{Ekstrom2015}.}
        \label{fig:Proton_Radii_sp_NNLOsat_plt}
\end{figure*}

Finally, Figs.~\ref{fig:Neutron_Radii_sp_DNNLOGO394_plt} to~\ref{fig:Proton_Radii_sp_NNLOsat_plt} show the single-particle rms radii similarly to Figs.~\ref{RMS_neutrons} and \ref{RMS_protons}, but the SCGF and IMSRG calculations labeled with (c) and (d) are now using the $\Delta$N$^{2}$LO$_{\rm GO}$\,(394)~\cite{Jiang2020} and the NNLO$_{\mathrm{sat}}$~\cite{Ekstrom2015} interactions instead of the 1.8/2.0~(EM7.5) interaction~\cite{Arthuis2024} for comparison.


\end{document}